\documentclass[aps,prb,twocolumn,showpacs,superscriptaddress]{revtex4-2}

\usepackage{amssymb}
\usepackage{amsbsy}
\usepackage{amsmath}
\usepackage{enumitem}
\usepackage[pdftex]{graphics}
\usepackage{color}
\usepackage{float}
\usepackage{comment}
\usepackage{graphicx}
\usepackage{dcolumn}
\usepackage{subfigure}
\usepackage{bm}
\usepackage{natbib}
\def\beq{\begin{equation}}
\def\eeq{\end{equation}}

\def\ra{\rangle}
\def\la{\langle}

\usepackage{bm}        
\usepackage{amssymb}   
\usepackage{exscale}
\usepackage{xcolor}

\newcommand\bea{\begin{eqnarray}}
\newcommand\eea{\end{eqnarray}}

\newcommand{\noi}{\noindent}
\newcommand{\non}{\nonumber}
\newcommand{\al}{\alpha}

\newcommand{\be}{\beta}

\newcommand{\ga}{\gamma}

\newcommand{\six}{\sigma_{i}^{x}}

\newcommand{\sjx}{\sigma_{j}^{x}}

\newcommand{\aal}{a_{\alpha}}

\newcommand{\xal}{|x_{\alpha}\rangle}
\newcommand{\xbet}{|x_{\beta}\rangle}
\newcommand{\balb}{b^{\alpha}_{\beta}}
\usepackage{amsmath}

\definecolor{darkred}{rgb}{0.90,0.2,0.2}

\usepackage[plainpages=false,pdfpagelabels,colorlinks=true,linkcolor=red,urlcolor=blue,citecolor=blue,pdftitle={},pdfauthor={},pdfdisplaydoctitle={},pdfduplex=DuplexFlipLongEdge]{hyperref}

\newcommand{\SzAna}{\langle\sigma^{z}(t)\rangle_{Analytic}}

\DeclareMathOperator{\sech}{sech}

\begin{document}

\title{Slow Dynamics and Kohlrausch Relaxation in Isolated Disordered Many-Body Systems}

\author{Asmi Haldar}

\affiliation{Max Planck Institute for the Physics of Complex Systems, N{\"o}thnitzer Stra{\ss}e 38, D-01187, Dresden, Germany}

\begin{abstract}
The Kohlrausch(-Williams-Watts) law of stretched exponential 
relaxation has been  observed for more than a century and a half in diverse complex classical systems. 
Here we  show that this law describes relaxation quite generically in closed (executing Schr\"{o}dinger dynamics), interacting disordered many-body systems across a range of system sizes using interaction range and disorder strength as primary tuning parameters.
This we observe for both time-independent and periodically driven (Floquet) systems. 
Finite-size analysis  indicates  the persistence of this non-thermal relaxation regime in the thermodynamic limit
thus defining a distinct dynamical regime.
This regime exhibits a peak in the time-scale
of the perceptible relaxation, upon crossing over from weak to strong disorder. We provide a simple  picture of this behavior, which naturally accounts for its general occurrence. 
Formation of spin-glass  -- one of the possible mechanisms for 
stretched relaxation appears incidental to the occurrence of Kohlrausch law in our context. 
Finally, we provide a simple non-Hermitian Hamiltonian formulation for the dynamics of a single spin embedded in the disordered chain. This provides an analytical formula that captures not only the Kohlrausch relaxation of the disorder averaged auto-correlation but also captures the largely diverse dynamics of
an arbitrary target spin in the system. Our work hence also provides a concrete quantification of the ``pre-thermal slowness" in many-body
disordered system.

\end{abstract}

    \date{\today}
\maketitle

\section{Introduction}
\label{Intro}
Strong Disorder is believed to result in absolute localization in many-particle systems even in the presence of interactions - a phenomenon known as many-body localization (MBL)~\cite{MBL_Altshuler,MBL_Gornyi,Huse_Pal_MBL,Huse_Oganesyan_MBL,Imbrie_MBL_Review,Abanin_MBL_Review}. 
MBL is believed to persist even in the presence of 
periodic drive and can exhibit
interesting new phases, collectively known as Floquet MBL phases~\cite{FlqMBL1,FlqMBL2}. A striking example of Floquet MBL phase is the discrete time crystalline (DTC) phase where the dynamics of a periodically driven system breaks the discrete time-translational symmetry of the time-periodic Hamiltonian
(see e.g., \cite{Brief_History,Else_Norm_DTC_Review,Asmi_Flq_Rev} and references therein).
In general, the actual stability of an MBL (and hence a DTC) phase in the thermodynamic limit is a matter of considerable debate~\cite{Community_for_MBL_1,Community_for_MBL_2,Prosen_Suntaj_NoMBL_1,Prosen_Suntaj_NoMBL_2,Jesko_NoMBL,Anatoli_Sels_NoMBL,Morningstar_NoMBL,Arti_MBL_Exists_2023}, though the stability of a DTC over experimentally relevant length and time-scale has been established already~\cite{Lukin_DTC,Monroe_DTC,Prethermal_Without_T,NISQ_2,Google_DTC}. \\

Here, we leave this debate to the side and address a different aspect of a Many-Body Disordered  (MBD) system -- what, if any qualitative difference in the relaxation behaviour of local correlations occur as a consequence of disorder. Our results not only provide a positive answer to this question, but also uncovers a connection 
between the non-thermal dynamics of an MBD system, and a broadly occurring yet somewhat enigmatic law that has been observed in the relaxation of complex systems in the classical world over more than a century and a half. \\

The relaxation dynamics of a many-body system may slow down qualitatively due to disorder and inhomogeneity. In the classical regime, the mechanism of slowing down
are believed to be quite different in different cases, but a surprisingly general law
of relaxation applies under remarkably diverse circumstances. This is known as the famous Kohlrausch law of stretched exponential relaxation~\cite{Kohlrausch}, where 
\begin{equation}
    {\rm Auto~Correlation} ~ \sim \exp{\left\{-\left(\frac{t}{\tau}\right)^{\gamma}\right\}},
\end{equation}
\label{Kolrausch_Law}
\noindent
where $0<\gamma\le 1,$ with $\gamma = 1$ giving the simple exponential relaxation.
Examples include slow dynamics of disordered classical systems like structural glasses and super-cooled liquids~\cite{StrchdExp_MolElec_Glasses_Review}, spin glasses~\cite{Binder_Young_RMP} and visco-elastic media~\cite{Kohlrausch}.
The origin and ubiquity of the Kohlrausch law has remained an intriguing open problem of physics and chemistry in spite of powerful attempts to address it (see, e.g.,~\cite{Palmer_Anderson_StretchedExp, Spohn_Kohl}). Here we show that the Kohlrausch law also appears in the zero-temperature Schr\"{o}dinger dynamics of interacting quantum disordered many-body systems. The Kohlrausch form with its
stretching exponent ($\gamma$) and the time-scale ($\tau$) provides a quantification of the well-known ``slowness" of heating in the so called ``prethermal regime"~\cite{Kuwahara_Mori_Saito_Prethermal,Mori_Kuwahara_Saito_Prethermal,Dima_Prethermal,Dima_Floquet_Prethermalization} in the context of localized quantum many-body  systems~\cite{Prethermal_Without_T,Else_Prethermal,Google_DTC}.
\begin{center}
\end{center}
\begin{figure*}[ht!]
\begin{center}
\includegraphics[width=0.3293\linewidth]{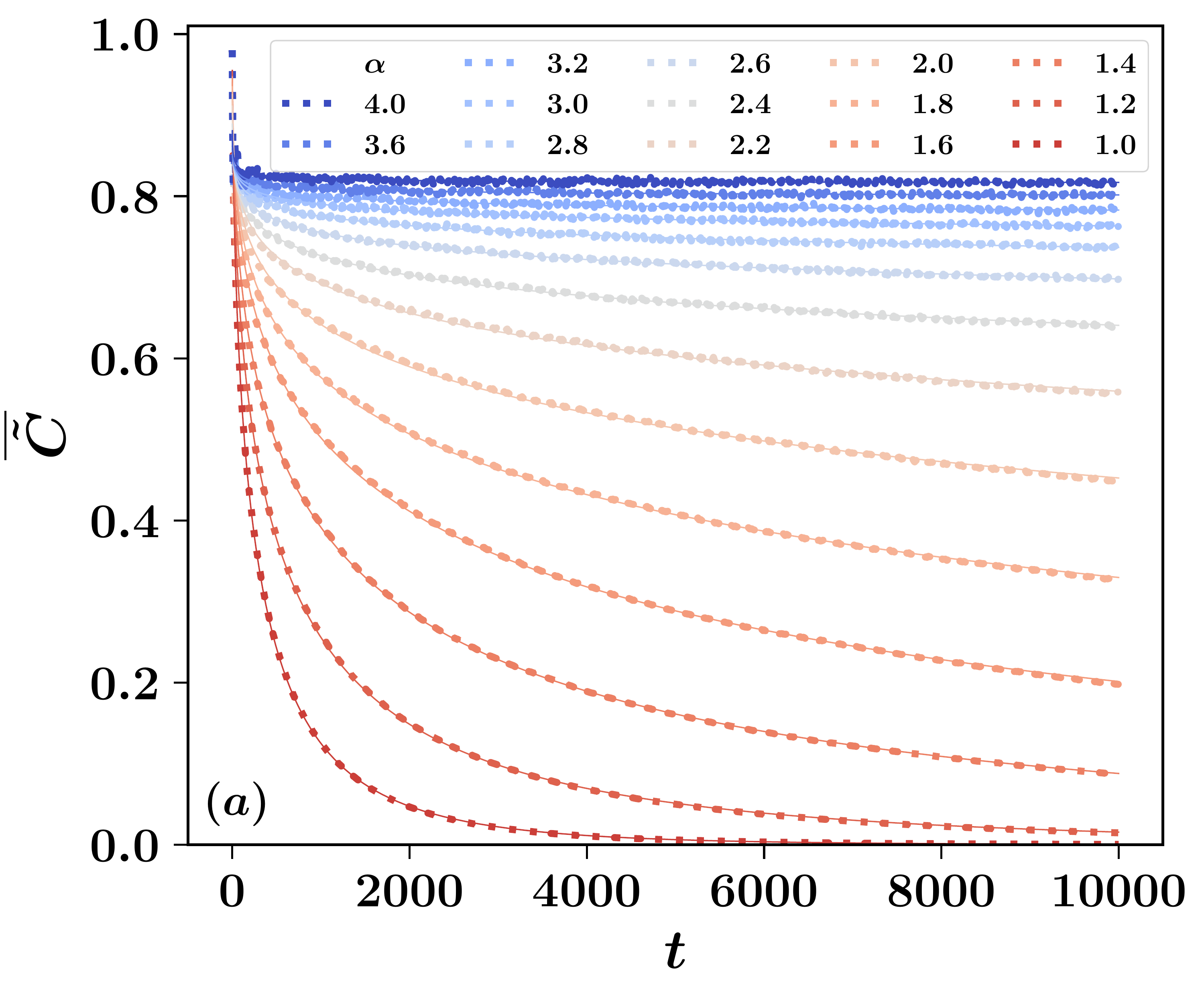}
\includegraphics[width=0.3293\linewidth]{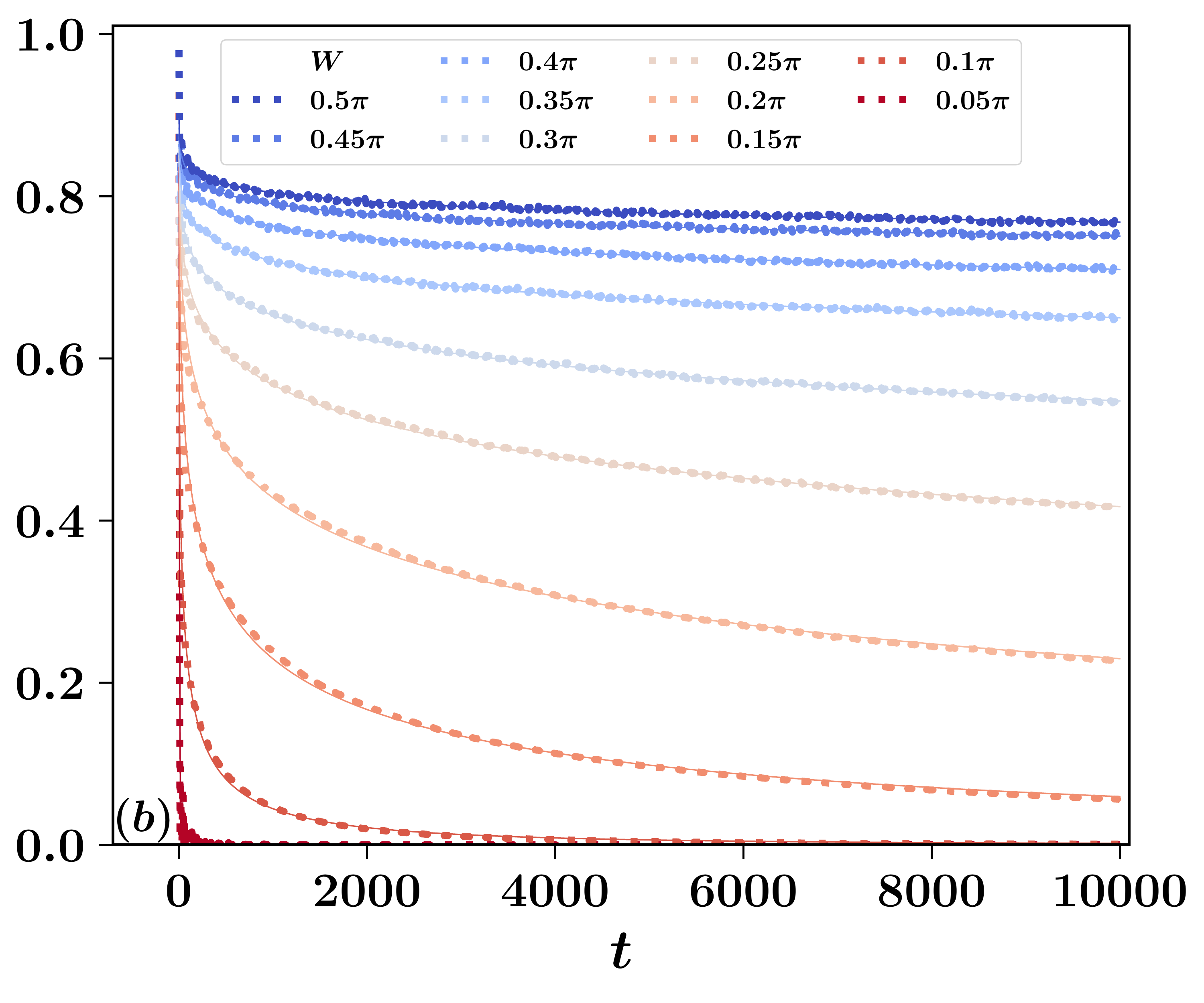}
\includegraphics[width=0.3293\linewidth]{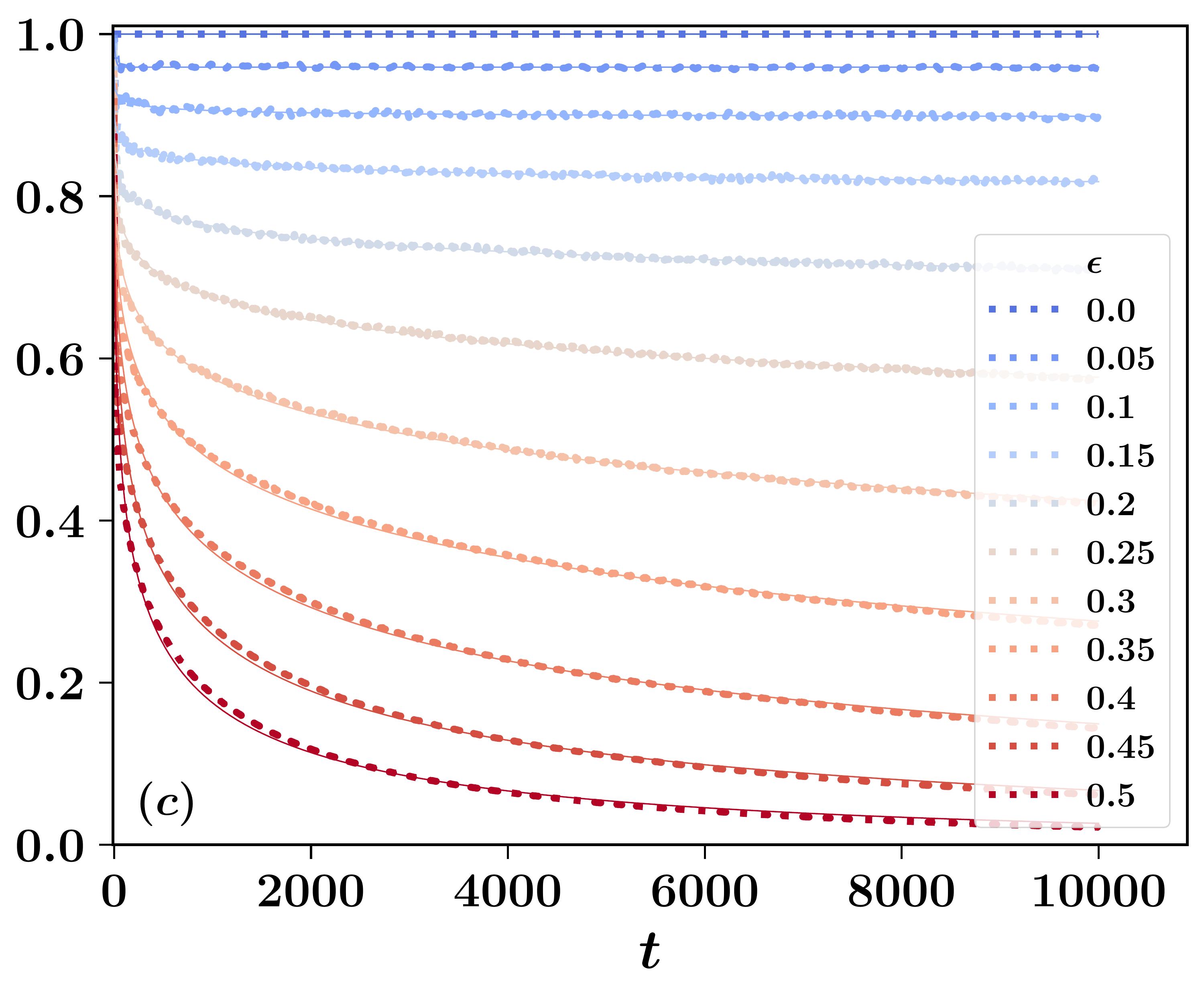}
\end{center}
\caption{Kohlrausch law of stretched exponential relaxation 
of the disorder-averaged auto-correlation $\overline{\tilde{C}}$
from least (bottom) to most localized regime (top) for $L=26$. Representative cuts across the multidimensional parameter space of the Floquet system (Eq.~\ref{Eq:FlqUnitary}) are shown (several others taken -- all show similar behaviour; data not shown).
Dotted lines are numerical data and the solid lines are the fittings of the data with a stretched exponential function 
(Eq.~\ref{Eq:Fit_Function}). From Left to Right, as a function of \textbf{(a)} $\alpha$ (keeping $W=0.4\pi$, $\epsilon =0.2$ ), \textbf{(b)} $W$ (keeping $\alpha=2.6, \epsilon=0.2$)and \textbf{(c)} $\epsilon$ (keeping $W=0.4\pi, \alpha=2.6$), respectively.}
\label{Fig1:Kohlrausch}
\end{figure*}
We demonstrate the occurrence of the Kohlrausch law in largely diverse settings - both with long and short range interactions, periodically driven (Floquet) Hamiltonians and static ones. For concreteness, we primarily focus on periodically driven quantum spin chains with power-law interactions, whose range can be tuned by varying the exponent. A short-range version of the model exhibits a stable DTC-MBL phase in appropriate parameter regimes in finite systems~\cite{Vedika_ACL_RM_SLS_DTC}. For this system we characterize the phenomenological features of the slow regime. Then we demonstrate those in other settings.
We numerically uncover distinct early and late dynamical regimes, and in none of those we find any significant signature of 
the putative dynamical phase of ``critical DTC"~\cite{Dima_CDTC,Else_Norm_DTC_Review} 
with scale-free relaxation in our one-dimensional geometry. 
\\

Finally, we provide an analytical formulation
based on a non-hermitian Hamiltonian approach to capture
the dynamics of an arbitrary target spin embedded in the disordered system. The idea is to describe the localization behaviour of the system
in terms of its action as a bath on an arbitrary spin embedded in it.
While the approach is incapable of
capturing multi-spin correlations, it captures the
single-spin dynamics sufficiently accurately, and hence
reproduces Kohlrausch relaxation of the disorder-averaged auto-correlation function with 
excellent accuracy. \\


The plan of the paper is as follows. 
In Sec~\ref{SecII:StretchedExponential} we demonstrate the Kohlrausch law in a power-law interacting Floquet setting, where the degree of localization is controlled by tuning various parameters. We also illustrate the salient phenomenological aspects of the Kohlrausch regime, especially the occurrence of slowest perceptible dynamics (a ``peak" in the relaxation timescale) between the strongly and weakly localized regimes. In Sec.~\ref{SecIII:Variants} we demonstrate the phenomenology for next nearest-neighbour interacting systems --
both Floquet DTC-type and  static ones. In Sec.~\ref{SecIV:PhysicalPic} we provide a
simple physical picture explaining the occurrence of the peak. 
The picture is based on identifying roughly three different relaxation
regimes separated by their characteristic time-scales.
In Sec.\ref{subsec:Non_Hermitian}, we resort to a non-hermitian
Hamiltonian description of a single spin in the system. For this we
provide an analytical expression for the dynamics of the 
single-spin auto-correlation function, which, upon disorder averaging,
exhibits the Kohlrausch relaxation with remarkable accuracy.
In Sec.~\ref{SecV:NoCDTC} we investigate the occurrence of the putative critical DTC phase in one spatial dimension. 
\section {The Kohlrausch Relaxation in Long-Range Interacting driven MBD System}
\label{SecII:StretchedExponential}

\subsection{Model and Observables}
\begin{figure*}[ht!]
\begin{center}
   \includegraphics[width=0.325\linewidth]{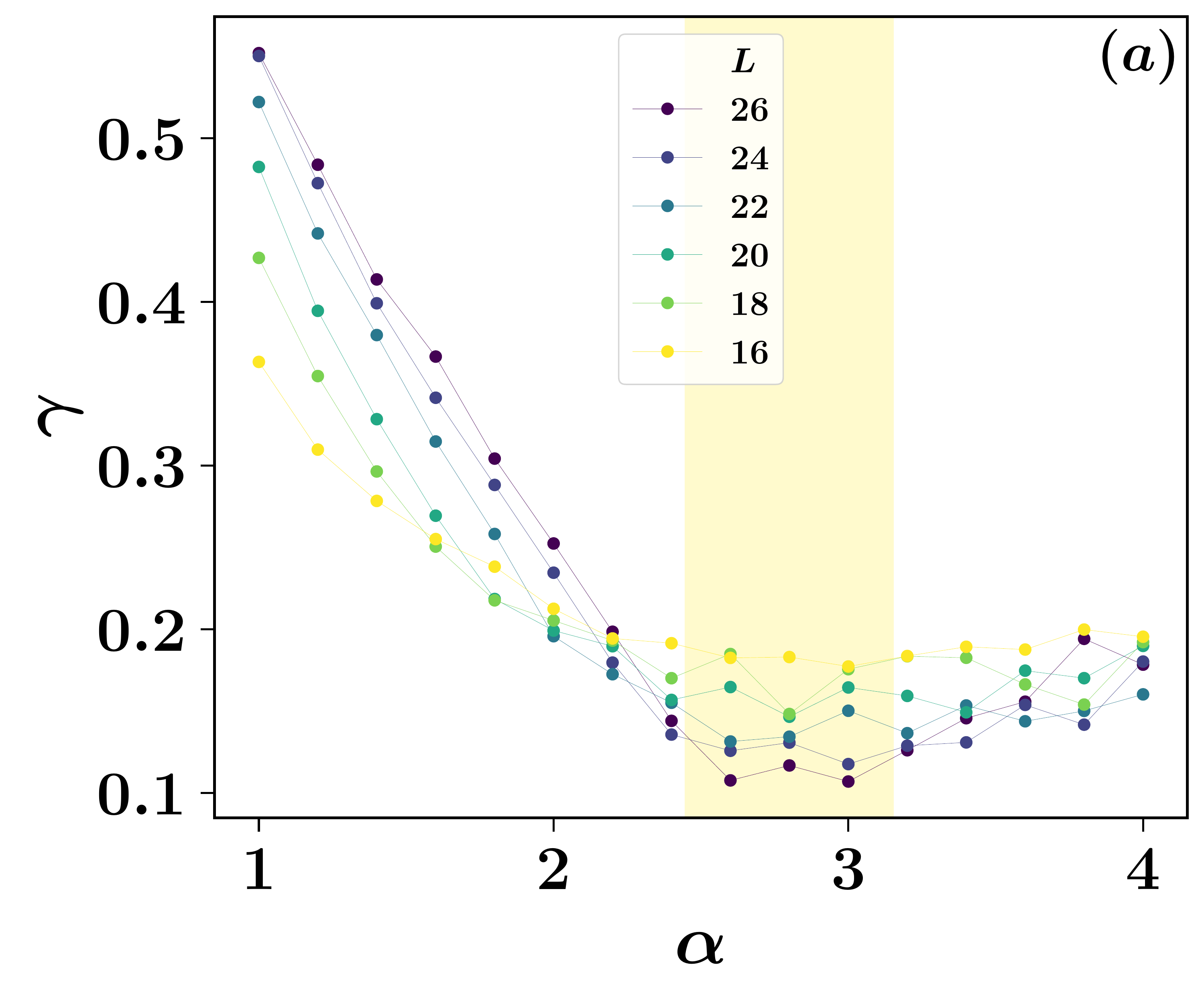}
    \includegraphics[width=0.325\linewidth]{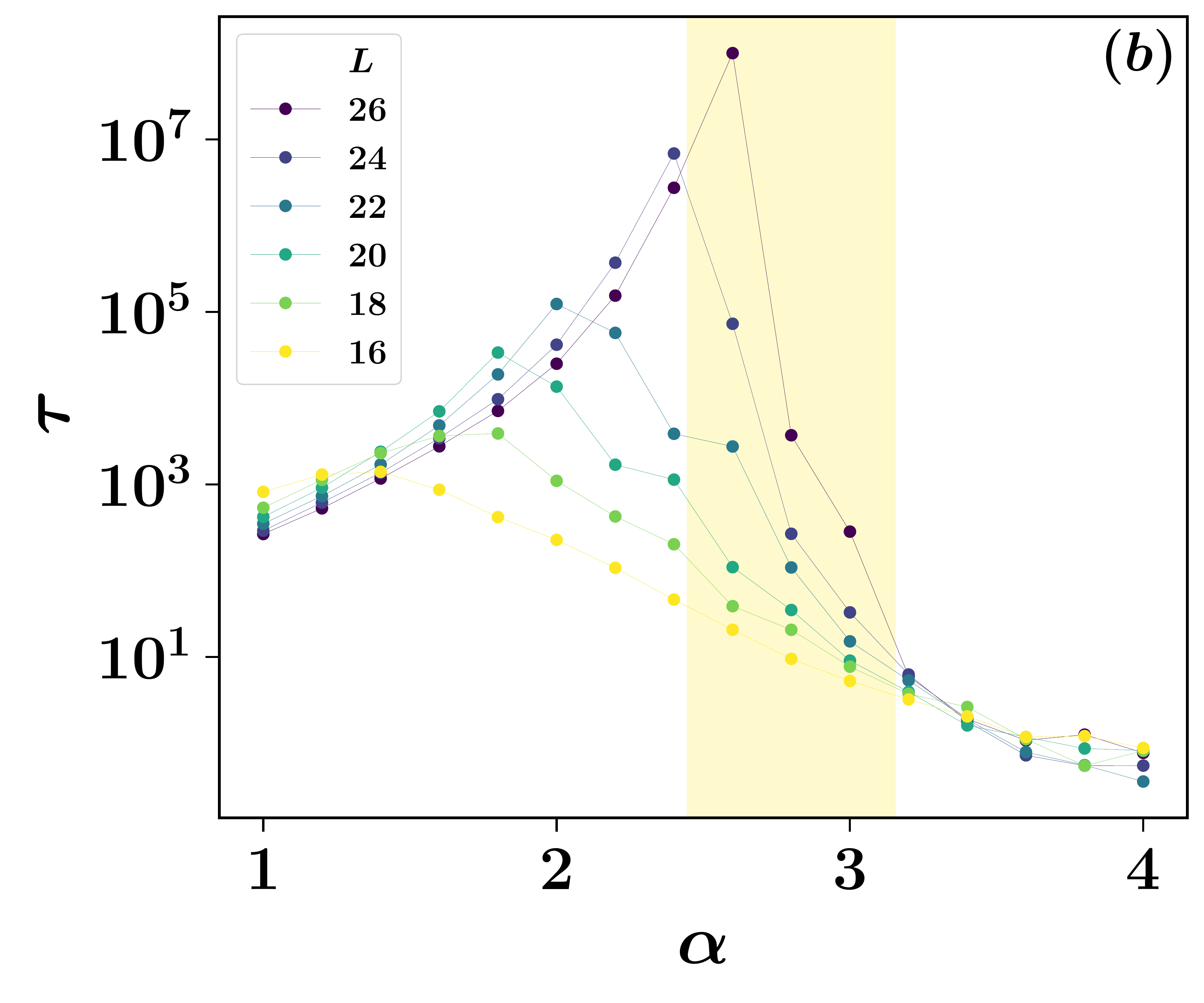}
    \includegraphics[width=0.325\linewidth]{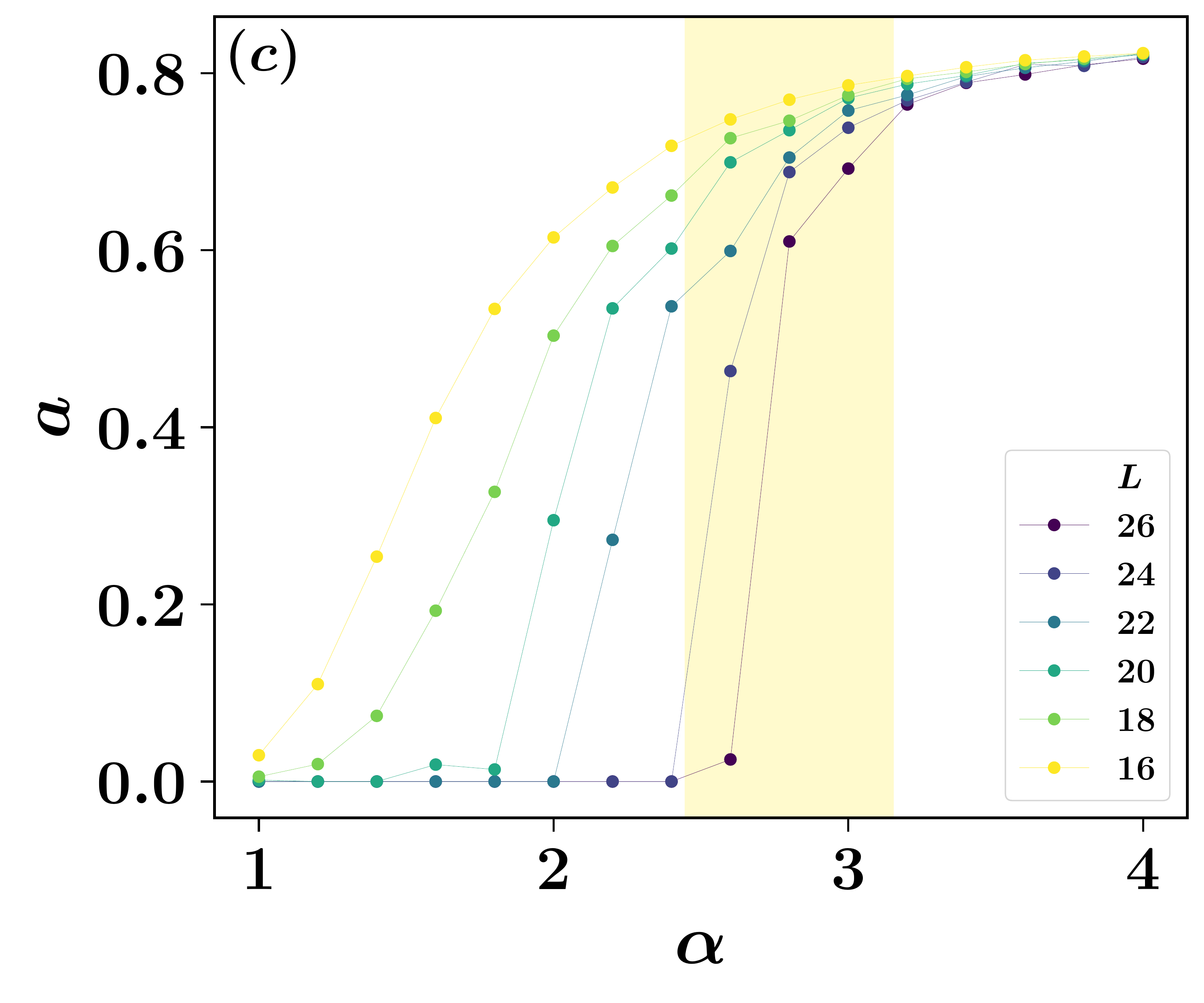}
    \includegraphics[width=0.325\linewidth]{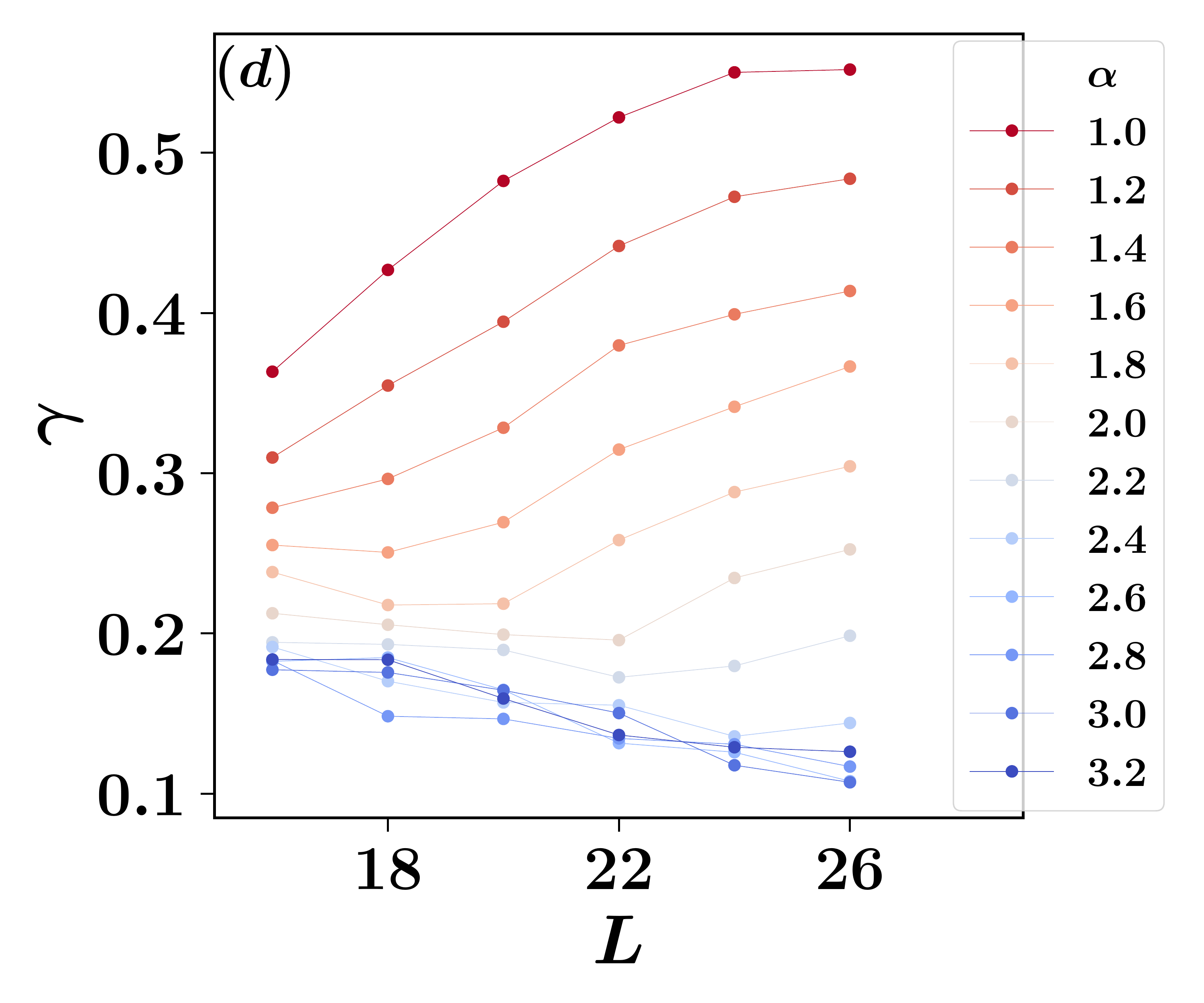}
    \includegraphics[width=0.325\linewidth]{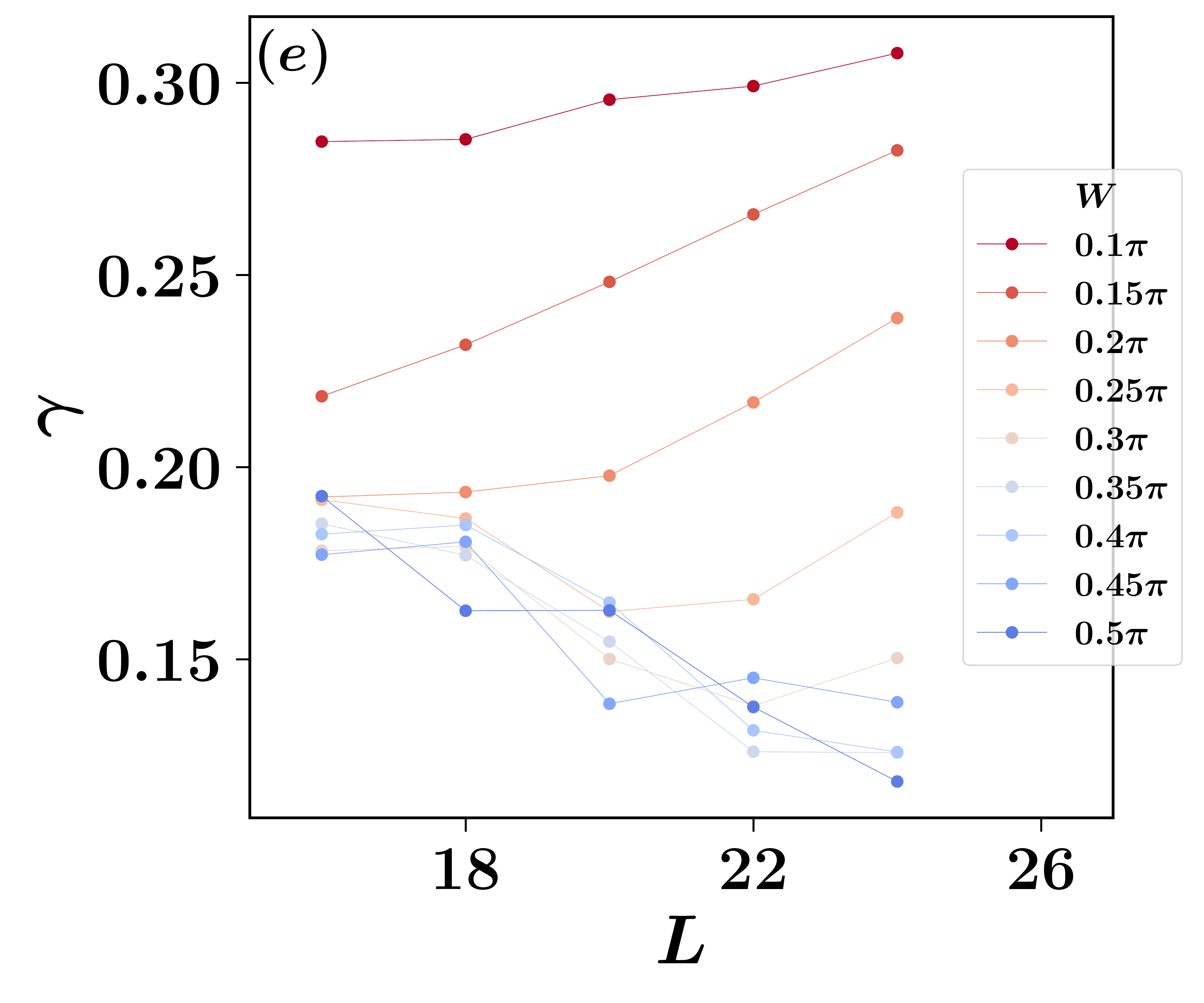}
    \includegraphics[width=0.325\linewidth]{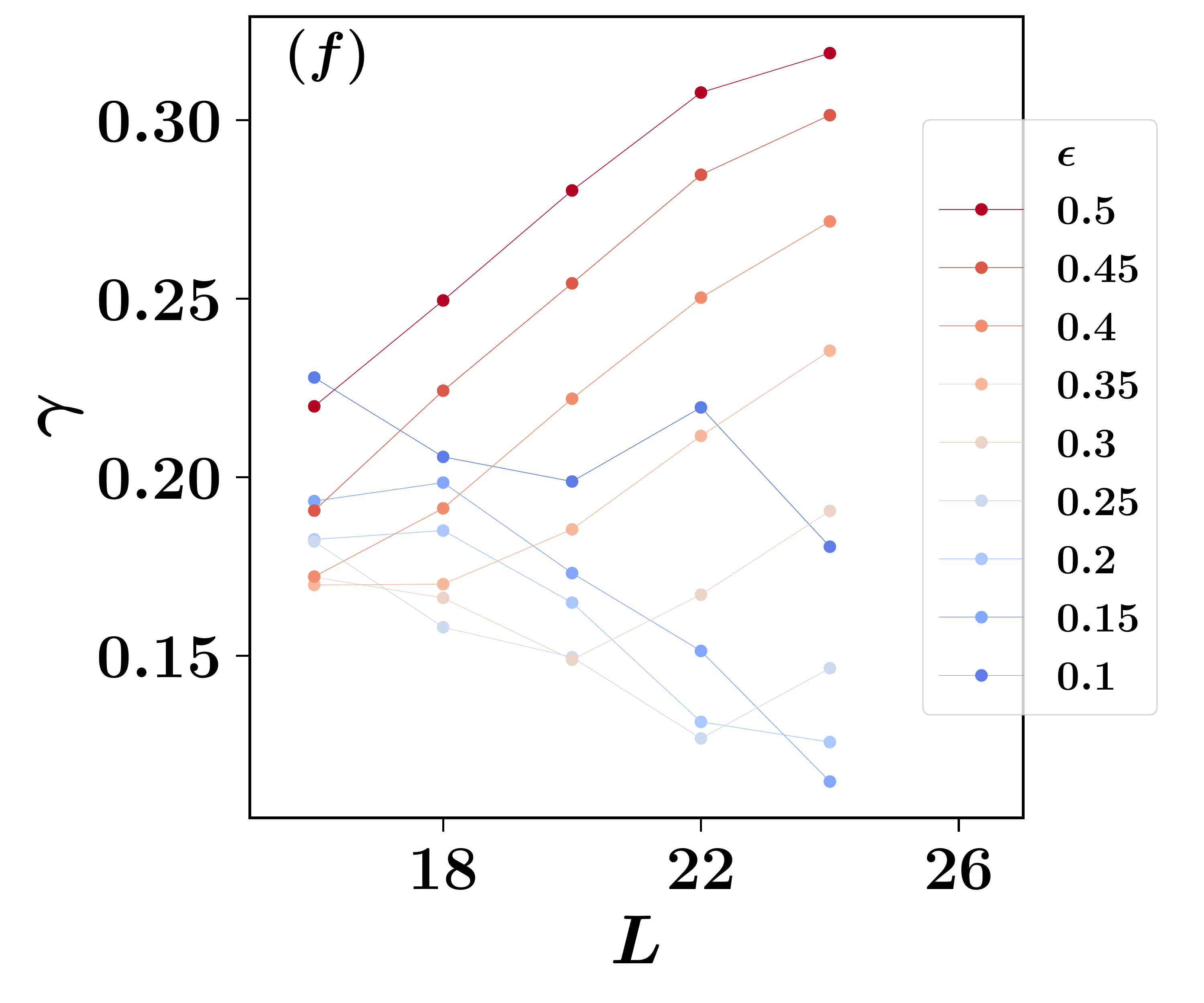}
\end{center}
  \caption{
  The Kohlrausch regime and its phenomenology:
  {\bf (a)} Behaviour of the stretching exponent $\gamma$ as a
  function of the exponent $\alpha$ of the power-law interaction (fixing $W=0.4\pi$, $\epsilon=0.2$). In the parameter regime marked with yellow, $\gamma$ shows systematic deviation from unity as $L$ is increased indicating stability of Kohlrausch law with increasing $L$ in that region. {\bf (b)}  Perceptible relaxation time-scale 
  $\tau$ vs $\alpha$ for various $L.$ A clear peak is
  observed in the intermediate values of $\alpha$. {\bf (c)} The residual auto-correlation, $a$ (see discussion after Eq.~\ref{Eq:Fit_Function}), as $t\to\infty$  obtained from evolution up to a finite time $T_{obs} = 10^{4}$). {\bf Lower Panel:} $\gamma$ vs $L$ for various variations of parameters:{\bf (d)} $\alpha$ (fixing $W=0.4\pi$, $\epsilon=0.2$) , {\bf (e)} $W$ (fixing $\alpha=2.6, \epsilon=0.2$)  and  {\bf (f)} $\epsilon$ (fixing $W=0.4\pi$, $\alpha=2.6$). In each frame, the curves in shades of red shows increase in $\gamma$ towards unity
   as $L$ is increased (instability of the Kohlrausch regime with increasing $L$), and those in shades of blue shows departure of $\gamma$ from unity with increasing $L,$ indicating
   stability of the Kohlrausch regime against increase in $L$.   }
  \label{Fig2:Phenomenology}
\end{figure*}
    \noindent
     In this section we consider the following Floquet-Unitary drive Hamiltonian in $1d$.
     \begin{equation}
	U_F = \exp{\left[-i \sum_{ij} \frac {J_{ij}}{r_{ij}^\alpha} X_i X_j\right]}
	\exp{\left[-i (\pi + \epsilon) \sum_i Z_i\right]}
	\label{Eq:FlqUnitary}
    \end{equation}
where $J_{ij}$ is chosen uniformly from $[0,W]$, and $X_i, Z_i$ are Pauli operators on site $i.$ A cycle consists of the application of this
unitary once on the wave-function.
The evolution within a cycle can be viewed as an evolution over a period
$T = 1 + \phi,$ where $\phi = (\pi +\epsilon)$ under the following piece-wise local time-periodic Hamiltonian.
\bea
H(t) = \nonumber \\
  &~& H_{z} = \sum_{i}Z_i, ~~~ nT \le t < nT + \phi \nonumber \\
   &~&  H_{x} =  \sum_{ij} \frac {J_{ij}}{r_{ij}^\alpha} X_i X_j, \nonumber \\
    &~& ~~~~~~~~~~~~~~~~~~ nT + \phi \le t <  (n+1)T. 
\label{Eq:FlqHam}
\eea 
\noindent
where $n$ are positive integers including $0.$\\

The long-range interactions above can be realizable and often even natural in experimental setups like ~\cite{Lukin_DTC, Monroe_DTC, Google_DTC, LR_Exprmnt_1,LR_Exprmnt_2,LR_Exprmnt_3,LR_Exprmnt_4,LR_Exprmnt_5}. 
Those setups are likely candidates for implementing
variants of the Hamiltonian given in Eq.~(\ref{Eq:FlqHam}). \\

The numerically exact simulation of the Floquet dynamics of the above form can be carried out by efficiently employing the Hadamard transform, reaching system sizes larger than those accessible via exact diagonalization (see, e.g. ~\cite{Tomaz_Hadamard}).
    Here we consider the (site-resolved) auto-correlation function defined for spin $i$ as 
    \begin{eqnarray}
    C_{auto}^{i}(nT) &=& \langle X_i(0) X_i(t) \rangle, ~t=nT \\
    \label{Eq:Ci}
    C_{auto}(nT) &=& \frac {1}{L}\sum_i C_{auto}^{i}, 
    \label{Eq:Def:Autocorr}
    \end{eqnarray}
    \noindent
    where $X_{i}(t)$ is the operator $X_{i}$ after $n$ complete cycles in the Heisenberg picture. \\
    
    For $\epsilon =0,$ $C_{auto}$ is trivially frozen to unity.
    For small but finite $\epsilon$ and large enough $W$ the above Hamiltonian has been argued to exhibit a DTC-MBL order for a short (effective) interaction range, i.e. large interaction exponent $\alpha$ (see, e.g.~\cite{Brief_History,Else_Norm_DTC_Review}).
    In a DTC, the sign of the auto-correlator flips every period, and this sub-harmonic oscillation persists in spite of the fact that the system is interacting and non-integrable. This happens for generic initial states, since each Floquet
    eigenstate is argued to be perturbatively close 
    (in $\epsilon$) to a cat state formed by
    a simultaneous eigenstate of $\{\sigma_{i}^{x}\}_{i=1,L}$ and its spin-flipped partner.  It is known from numerical studies that 
    regardless of whether there is an underlying stable MBL-DTC phase, the initial relaxation in an MBD system can slow down due to the disorders 
    (see, e.g.~\cite{Lezama,Lezama_LeaSantos_Slow_MBD}). 
%
To capture this slowness, we track the dynamics of the time staggered DTC order after each cycle, given by
    \begin{equation}
    \begin{split}
    \tilde{C^{i}}(nT) = (-1)^{(n+1)}C^{i}_{auto}(nT) 
                              ~ {\rm and} \\
        \tilde{C}(nT) = (-1)^{(n+1)}C_{auto}(nT).
        \end{split}
        \label{Eq:Def:Staggered_Autocorr}
    \end{equation}
    The Kohlrausch relaxation is manifested in the disorder-averaged values of the above quantities denoted by
    $\overline{\tilde{C^{i}}(nT)}$ and $\overline{\tilde{C}(nT)}$ respectively. The overhead bar denotes the
    averaging over disorder realizations as well as over various random $x-$bit-string initial states.\\
    
    
\noindent   
{\bf The Kohlrausch form for MBD Systems:} The presence of disorder in our system necessitates a
    formally slight but crucial modification of the standard form of the Kohlrausch law of stretched exponential form (Eq.~\ref{Kolrausch_Law}): we need to add a constant to it (e.g., to take care of a stable
    localization within a finite system-size or a finite observation time). This gives
    us the following form. 
    \begin{equation}
       y(t) = a + (1-a)\exp{\left\{-\left(\frac{t}{\tau}\right)^{\gamma}\right\}}.
        \label{Eq:Fit_Function}
    \end{equation}
    \noindent The added constant $a$ accounts for the residual (frozen) correlation as can be estimated from
    the data measured over a finite interval. This might or might not be the true residual auto-correlation within our resolution as $t\to\infty,$ depending on our measurement interval
    and/or the existence of a stable MBL phase (see the end of Sub-Sec.~\ref{SubsubSec:Relax_Params} for further discussions on this). \\
    
    %
    
    \subsection{Main Numerical Results}
    \label{SubSec:Main_Results}
    Here we summarize the main numerical results.
    In our system, the degree of localization can be tuned by three parameters. It can be increased by increasing the disorder strength $W,$ while can be decreased either by increasing $\epsilon$ (which
    enhances delocalization over the $x-$bit-string states) or increasing the effective range of interaction by decreasing the exponent $\alpha.$
    Fig.~\ref{Fig1:Kohlrausch}, shows various cuts
    in the parameter space representing the salient features of the relaxation: the disorder-averaged real-time dynamics has been shown (dotted lines).  For all the cases, throughout the entire parameter regime, the numerical data is fitted (continuous lines) very accurately with the Kohlrausch law (Eq.~\ref{Eq:Fit_Function}): The largest absolute least square error for fitting is $< 10^{-5}.$ This fits much better than 
    the other standard relaxations forms, namely, exponential and power-laws with and without
    additive constants.\\
    
    In Fig.~\ref{Fig1:Kohlrausch}$(a)$, plots for various values of $\alpha$ are shown
    for a fixed value of $\epsilon$ and $W$. As $\alpha$ is decreased, the dynamics switches from
    a strongly localized DTC-MBL regime for our finite system (where $\tilde{C}$ is frozen close to its initial value) to the delocalized regime (where $\tilde{C}$ decays rapidly to zero). In $(b)$ and $(c)$, the parameter regimes are traversed (from stronger to weaker localization) by increasing $\epsilon$ and decreasing $W$ respectively, keeping other parameters fixed. 
    \\

    \subsubsection{Phenomenology in the light of Relaxation Parameters:}
    \label{SubsubSec:Relax_Params}
    
    The relaxation parameters ($\gamma, \tau$ and $a$) are extracted from the fitting of the Kohlrausch form
    (Eq.~\ref{Eq:Fit_Function}) to $\tilde{C}(t)$ measured up to a finite observation time $T_{obs}.$ The aim of the fitting is to provide a 
simple yet sufficiently accurate (leading order) description of a complicated 
relaxation dynamics. The exact dynamics, of course have, several finer details and also various dynamical regimes. Hence there 
is apriori no reason to expect that the entire relaxation should be well approximatable by a single simple function. The values of the best fitting parameters as well as the suitable fitting functions 
can thus depend upon choice of the temporal regime under analysis (i.e., on $T_{obs}$). 
In our case, though the parameters do exhibit such a dependence (see Fig.~\ref{Fig:Var_wth_Tobs}), the functional form, quite interestingly, does not -- Kohlrausch form always provides the most optimal (over the set of usual physical relaxation forms like exponential, power-law etc with or without an additive constant etc) 
and also a very accurate description in all regimes. Though we have fitted the whole regime
with a single set of parameters in most cases here, the separation between the early time and late time regimes are actually quite pronounced (the short, early time dynamics and the longer late time dynamics does not fit equally well with a single set of parameters). This has been addressed in further details in Sec.~\ref{SecV:NoCDTC}. In the following we discuss the phenomenology and the significance of the relaxation parameters. 
\\
    
    \noindent
    {\bf The Stretching Exponent $\gamma$ and the Stability of Kohlrausch Regime:}
    Though the Kohlrausch's stretched exponential relaxation does not probably look qualitatively 
    very different from the thermal exponential relaxation, a unified understanding of the stretching has still remained elusive and enigmatic. The efforts in this direction (models with finite density of long-lived metastable states, hierarchical constraints etc) indicate that a unified explanation would 
    require elements fundamentally different from those responsible for exponential relaxation~\cite{StrchdExp_MolElec_Glasses_Review}. Hence the fate of the stretching exponent $\gamma$ in the
    thermodynamic limit, and identification of those regime where it is non-trivial in that limit, are issues of fundamental importance. \\
    
    In order to investigate the stability of the stretched exponential relaxation against the increase in system-size $L$, we extracted the stretching exponent $\gamma$ for various values of $L$ from real-time dynamics data over $10^4$ cycles averaged over disordered configurations. The results are shown in Fig.~\ref{Fig2:Phenomenology}(a),(d),(e),(f). The Fig.~\ref{Fig2:Phenomenology}(a) shows that there is a regime of $\alpha$ (shaded with yellow) where the stretching exponent $\gamma$ actually decreases
    systematically away from unity with increasing system-size $L.$ 
    This is what we identify as the ``slow-regime" or ``Kohlrausch relaxation regime". Away from the slow regime, $\gamma$ increases steadily with $L,$ and as is intuitively believed (see, e.g.~\cite{Lezama}), $\gamma$ vs $L$ might have unity as its asymptote, implying a thermal relaxation behaviour as $L\to\infty$. Between these two regimes, the trend is
    non-monotonic and ambiguous in our finite system simulation, and it is difficult to determine whether the two phases are separated by a transition or a crossover. \\
    
    The three frames in the lower panel of Fig.~\ref{Fig2:Phenomenology} shows the variation of $\gamma$ with $L$ both within (in shades of blue) and outside (in shades of red) the Kolhrausch regime. In Fig.~\ref{Fig2:Phenomenology}$(d)$ , the results are shown for various values of $\alpha$ for a fixed value of $W$ and $\epsilon$ (corresponding to those
    in $(a)$). In Fig.~\ref{Fig2:Phenomenology}$(e)$ and $(f)$ the parameter space
    are traversed by varying $W$ and $\epsilon$ respectively.
    \\
    
    The possibility of the existence of an intermediate phase/regime between a thermal and the non-ergodic phase in a disordered many body system has been indicated long ago (see, e.g.~\cite{Luitz_Yevgeni_Review} for a review) and some of the dynamical consequences of such an anticipated regime have also been considered more recently~\cite{Luitz_Yevgeni_Nonergodic_Phase, Anatoli_Sels_Intermediate_Regime}. The Kohlrausch regime puts all those observations on a unified ground, yet it does not necessarily require a stable MBL phase as an asymptote. Strong enough disorder which gives the local integrals of motions~\cite{Abanin_MBL_Review, Imbrie_MBL_Review} a life-time $> T_{obs}$ is sufficient. \\
    
\begin{figure*}[ht]
\begin{center}
\includegraphics[width=0.32\linewidth]{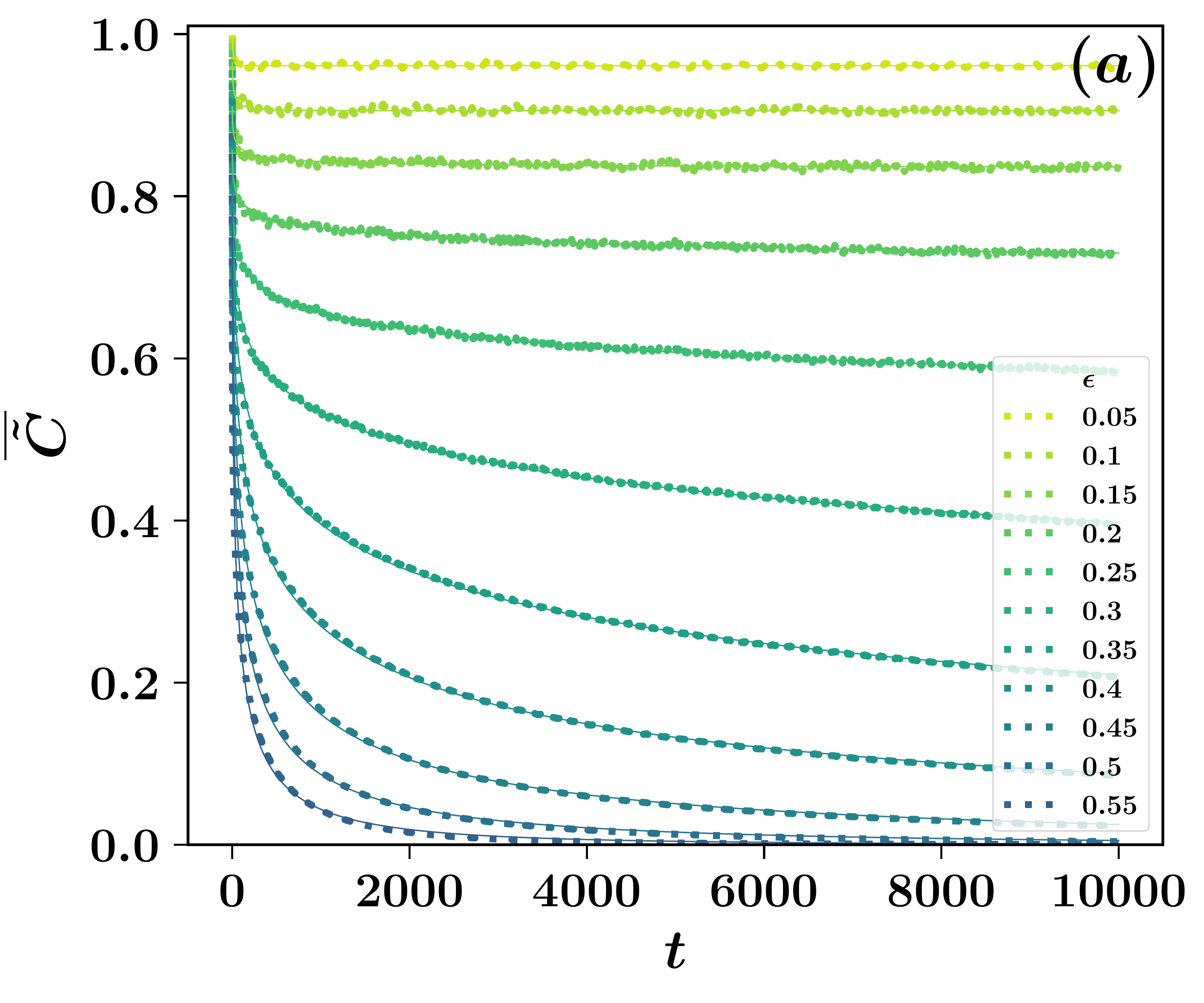}
\includegraphics[width=0.32\linewidth]{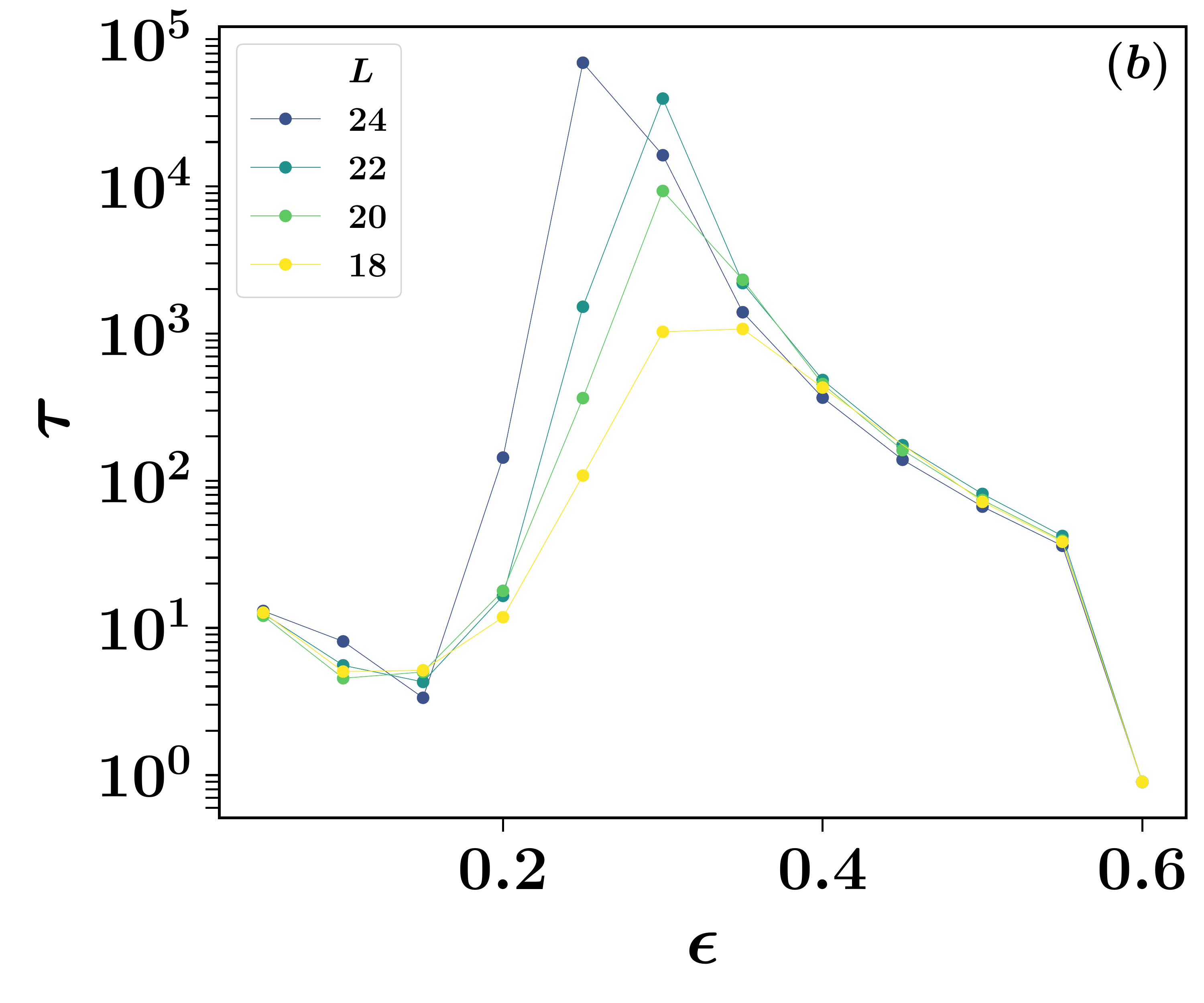}
\includegraphics[width=0.32\linewidth]{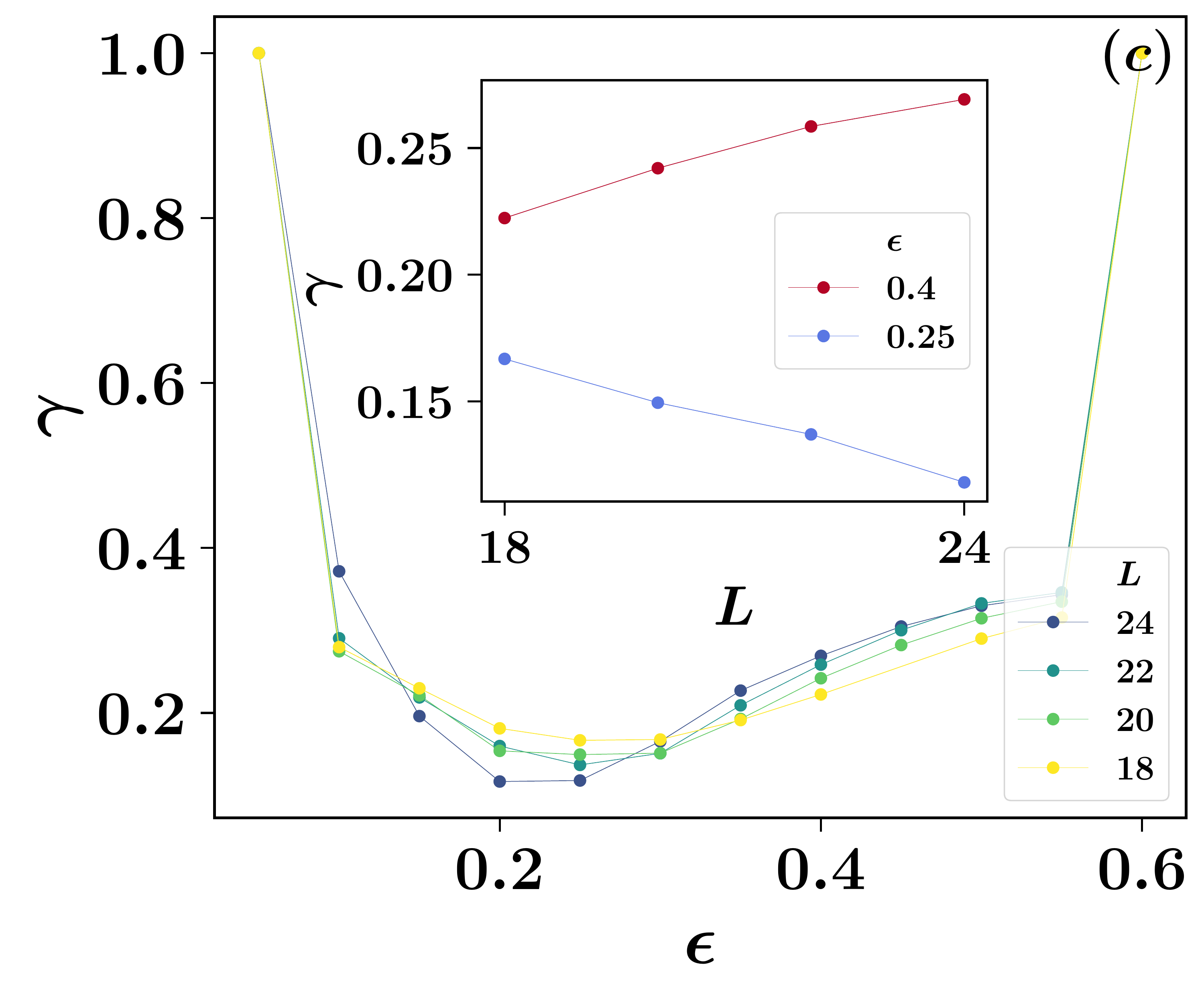}
\end{center}
\caption{Kohlrausch relaxation in up to next-nearest neighbour interacting Floquet system (Eq.~\ref{Eq:FlqUnitaryShort}). \textbf{(a)} Auto-correlator for various values of $\epsilon$, for a fixed disorder strength, for $L=24.$ Dotted lines are actual data from the numerics, and solid lines are fitting of the data with a stretched exponential function defined at Eq.~(\ref{Eq:Fit_Function}). \textbf{(b)} Time scale $\tau$ as a function of $\epsilon.$ \textbf{(c)} Stretching exponent $\gamma$ as a function of $\epsilon.$ The {\bf inset} shows the dependence of $\gamma$ on the system size for at various values of $\epsilon.$ The stable and unstable flows of Kohlrausch exponents with $L$ are shown respectively in blue ($\epsilon=0.25$) and red ($\epsilon=0.40$).  Similar behaviour can be observed as a function of the disorder strength for fixed values of $\epsilon$ (data not shown).
}
\label{Fig:ShortRange_Flq}
\end{figure*}

\begin{figure}[ht]
\begin{center}
\includegraphics[width=0.77\linewidth]{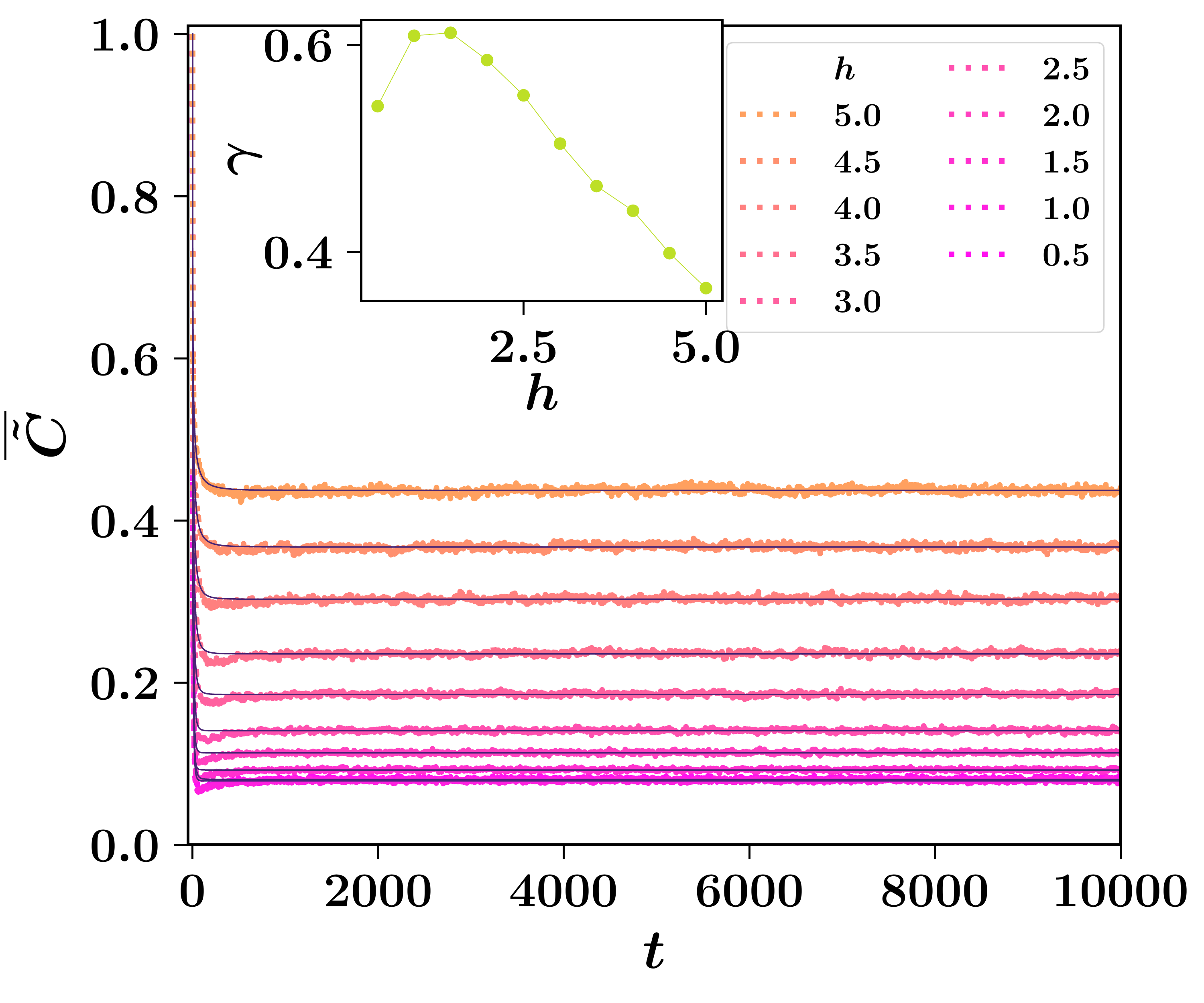}
\end{center}
\caption{
Stretched exponential relaxation for quench with a time-independent MBD
Hamiltonian (Eq.~\ref{Eq:Heisenberg}). The inset shows the variation of the
stretching exponent $\gamma$ with the disordered strength $h.$
}
\label{Fig:Heisenberg}
\end{figure}
    \noindent
    {\bf Peak in Relaxation Time-scale $\tau$:}  Next we focus on another 
    striking aspect of our phenomenology. The relaxation time-scale $\tau$ does not monotonically increase
    as one moves from a relatively weakly localized regime to a more strongly localized one. Instead, $\tau$
    exhibits a lofty peak somewhere between the two said regimes. This is shown in Fig.~\ref{Fig2:Phenomenology}(b). This might appear counter-intuitive at first,
    since it is sensible to associate slower dynamics (larger $\tau$) with a more strongly localized regime.
    However, for a finite $T_{obs}$ and a finite fitting resolution, there could be
    a substantial fraction of spin which would appear completely frozen, and our finite time observation will
    not sense their participation in the relaxation and club it in the constant $a$ (discussed in further details in the following). An explanation and interpretation of the peak and its behaviour with 
    the change in parameters have been given in terms of a simple physical picture based on a few very reasonable physical assumptions in Sec.~\ref{SecIV:PhysicalPic}. \\

    \noindent
    {\bf Interpretation of the Residual auto-correlation $a$:}
    The parameter $a$ provides a measure of the degree of localization as 
    extractible from the data for dynamics over a finite $T_{obs}.$ Our
    finite-time characterization of weak and strong localization 
    is based on the value of $a$ ($a = 0$ implies delocalization
    where $a = 1$ means perfectly frozen). However, it may be noted that
     $a$ is neither the residual auto-correlation 
after time $T_{obs}$ nor (necessarily) the remnant correlation observed after infinite time (i.e., for $T_{obs}\to\infty$).
It is the asymptotic auto-correlation {\it estimated} from the data obtained over a {\it finite} $T_{obs}.$ 
If there were only one dynamical regime that could be captured accurately by a single fitting function with a fixed set of parameter values, then of course everything would be independent of $T_{obs},$ and $a$ could be interpreted as the residual correlation after infinite time.
But as behooves a complicated many-body system, an MBD system exhibits various distinct stages of dynamics in the sense that a single function with a fixed set of parameter values cannot capture the dynamics over all time scales. Hence, $a$ just gives an estimate of the part of the correlation that 
looks completely frozen over an observation time-scale $T_{obs}$ within the accuracy with which the Kohlrausch description is expected to hold. 
Thus, a non-vanishing $a$ extracted by finite-size analysis from a finite $T_{obs}$ data does not necessarily imply existence of a stable MBL. 

\section{Short Range Variants: Floquet and Static MBDs}
\label{SecIII:Variants}

 \noindent
    In this section we consider another Floquet-Unitary drive for a strictly
    short range system with interactions ranging up to the next nearest neighbour on the same lattice geometry as before. The evolution operator
    over a period is given as follows.
     \bea
	U_F &=&
	\exp{\left[-i H_x \right]}
	\exp{\left[-i (\pi + \epsilon) H_z\right]} \nonumber \\
 where, ~ H_{z} &=& \sum_{i}Z_i, \nonumber \\
     H_{x} &=& - \sum_{i} J_{i} X_i X_{i+1} + J_2 \sum_{i} X_i X_{i+2}
	\label{Eq:FlqUnitaryShort}
    \eea
    $J_i$'s are chosen uniformly from $[0.1\pi,0.4\pi].$
    A static version
    of the model in a transverse field is believed to show a localization-delocalization transition ~\cite{Kjall_Frank} as the disorder strength is tuned. We choose a parameter regime where the localization is strong enough, though, like in the earlier case, is not based on existence of a stable MBL phase.\\
    
    The evolution shows qualitatively same phenomenology as obtained for the long-range system, as shown in Fig.~\ref{Fig:ShortRange_Flq}. Here the
    degree of localization is controlled by tuning the imperfection $\epsilon$
    ($\epsilon=0$ is trivially localized). The left frame shows real-time dynamics fitted with the stretched exponential form in Eq.~\ref{Eq:Fit_Function}. The middle panel show the peak in the relaxation time-scale $\tau$ as a function of $\epsilon.$ The right frame shows the value of the stretching
    exponent $\gamma$ as a function of $\epsilon$ for various values of $L.$
    A stable regime for Kohlrausch relaxation is clearly observed (where $\gamma$ decreases with increasing $L.$) The inset shows the variation
    of $\gamma$ with $L$ in (blue) and away (red) from the stable regime.

Next, we consider the anisotropic Heisenberg model ~\cite{Log_Entanglement}, in a disordered field, and study the relaxation dynamics of a state under this Hamiltonian:
\begin{equation}
    H = J \sum_{i} (X_i X_{i+1} + Y_i Y_{i+1}) + J_z  \sum_{i} Z_i Z_{i+1} + \sum_{i} h_i Z_i
	\label{Eq:Heisenberg}
\end{equation}
$h_i$ are chosen uniformly from $[h,-h]$, $J_z=0.5 J, J=1.$

\section{A Phenomenological Picture of the Kohlrausch Scenario 
and A Non-hermitian Analytical Approach}
\label{Sec:Non_Hermitian_Theory}

\subsection{A Simple Physical Picture}
\label{SecIV:PhysicalPic}
\begin{figure}[ht!]
\begin{center}
\includegraphics[width=0.97\linewidth]{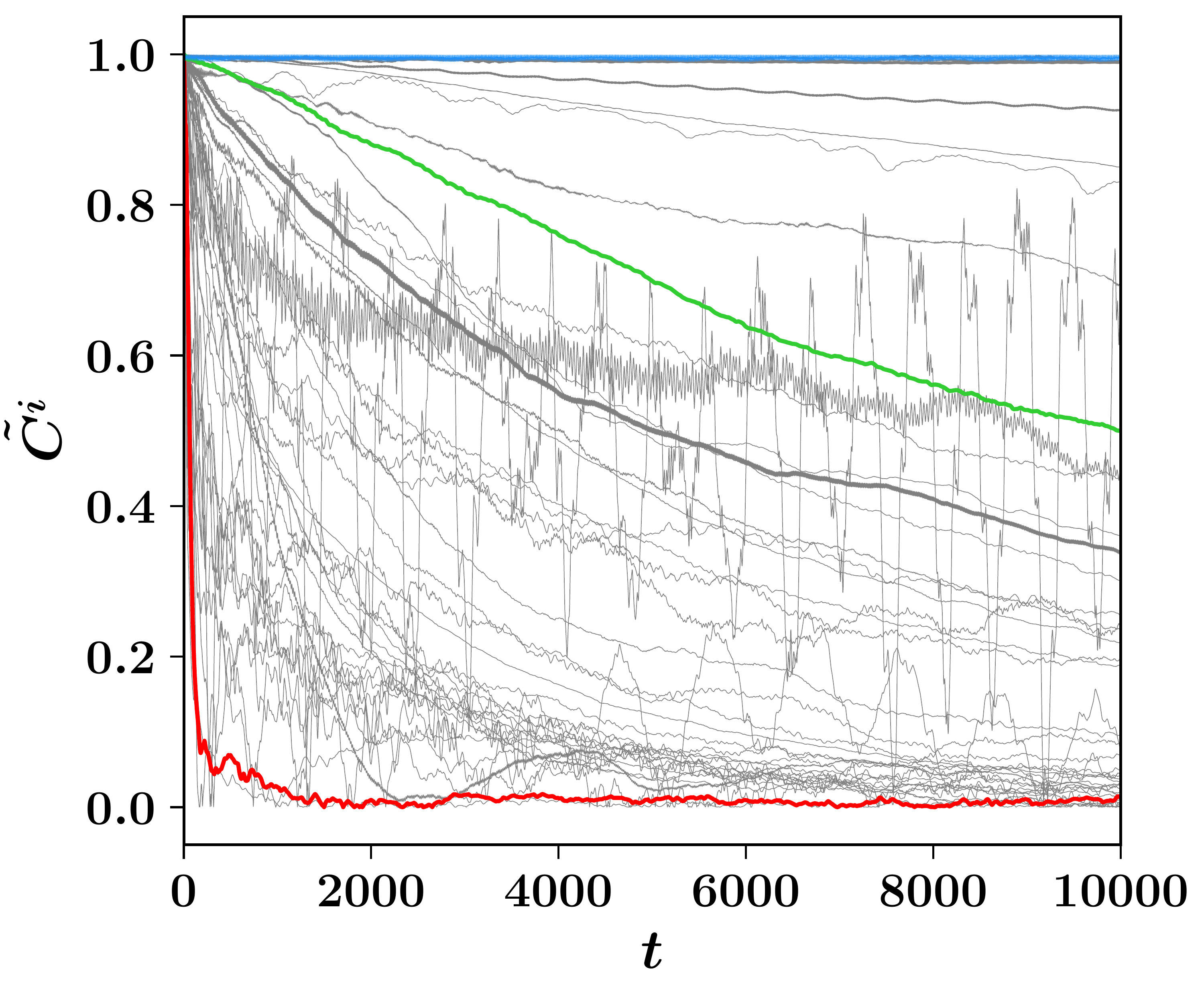}
\end{center}
\caption{
Real-time dynamics of single spins sampled completely randomly from various realizations, for $\alpha=2.5, W=0.4\pi, \epsilon=0.2.$ The plot shows a high degree of variability in the local dynamics with different types of behavior appearing with appreciable probabilities.  
}
\label{Fig:5A_SinglSpnDyn}
\end{figure}

In this section, we provide a cartoon of a qualitative picture underpinning the phenomenology entailing Kohlrausch relaxation described in Sec.~\ref{SecII:StretchedExponential}. A remarkable heterogeneity in the microscopic dynamics underpins the simple Kohlrausch law, as can be gleaned from a representative plot of the real-time dynamics of single spins chosen randomly from various samples (Fig.~\ref{Fig:5A_SinglSpnDyn}) for a set of parameter values for which the Kohlrausch law remains stable with increasing $L$. Here we provide clarifications of central aspects of the phenomenon: we explain salient features of the Kohlrausch phenomenology (e.g., the occurrence of the peak in $\tau,$ $L-$dependence of $\gamma,$ the behavior of the peak in the $T{obs}\to\infty$ as well as $L\to\infty$ limit) employing a simple physical picture.\\

\begin{figure*}[ht!]
\begin{center}
\hspace{1.2cm}
\includegraphics[width=0.29\linewidth]{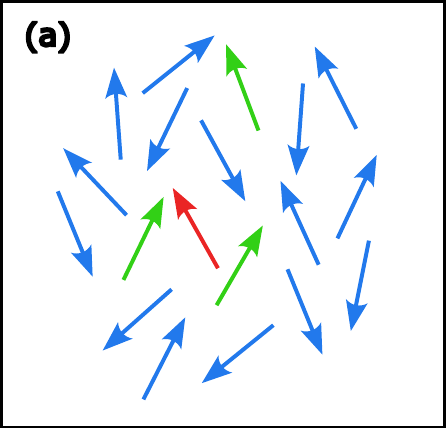}
\includegraphics[width=0.29\linewidth]{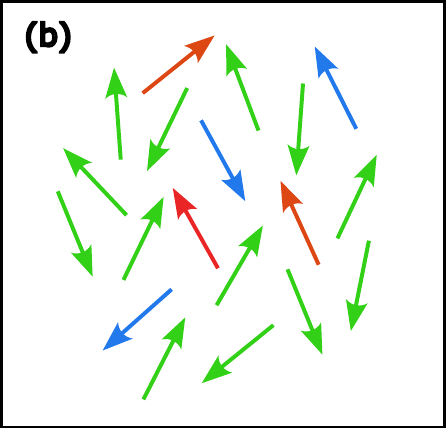}
\includegraphics[width=0.29\linewidth]{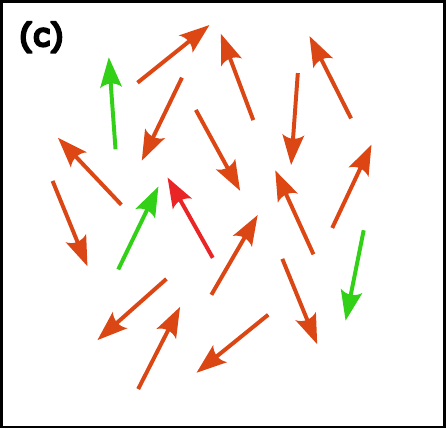}
\includegraphics[width=0.98\linewidth]{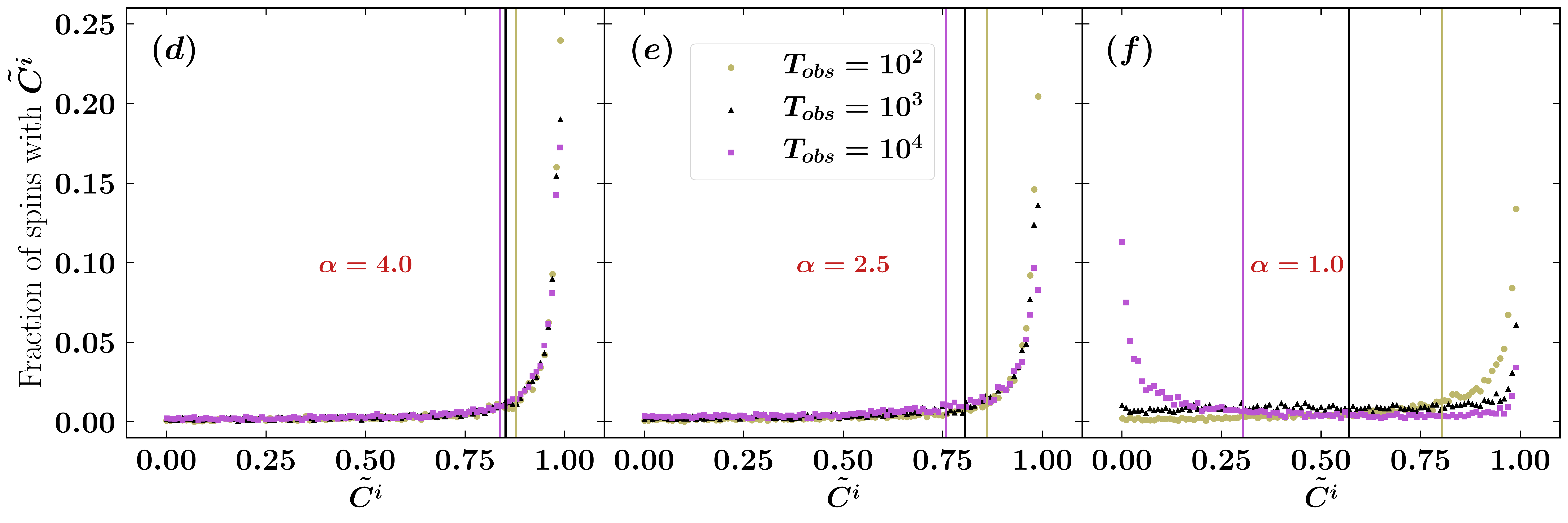}
\end{center}
\caption{The physical picture: {\bf Top:} A sequence of cartoons of three relaxation scenarios. \textbf{(a)} Strongly localized, dominated by frozen auto-correlation (blue arrows), \textbf{(b)} Kohlrausch regime, dominated by slowly decaying auto-correlation (green arrows), and  \textbf{(c)} mostly delocalized, dominated by rapidly relaxing auto-correlation (red arrows). \\
{\bf Bottom:} 
Histogram of single-spin auto-correlations  $C_{auto}^i$ (Eq.~\ref{Eq:Ci} ) sampled over an ensemble of randomly chosen 5000 spins, corresponding to the respective cartoon above. 
Three different colors show results after three successive decades of evolution. The mean single-spin auto-correlation of the ensemble at three different times is shown by the vertical lines.
}
\label{Fig5B:Physical_Picture}
\end{figure*}

\noindent
{\bf The Basic Ingredients: The Slow, The Fast, and The Frozen:} 
Our physical picture is based on a cartoon (Fig.~\ref{Fig5B:Physical_Picture} (a)-(c)). The observation window $T_{obs}$ defines the reference timescale. Though there is a broad distribution of relaxation times, the cartoon is built on three variants in particular: spins that relax fast 
(red arrows in the cartoon), slow spins -- whose relaxation is tiny but discernible during the process of the observation (green arrows), and spins whose relaxation (if any) within $T_{obs}$ is beyond our 
resolution for detecting any dynamics (blue arrows). 
The dynamics can be broadly classified into three regimes (as described below), and its features can be qualitatively captured in terms of the dynamics of these cartoon spins. The relative populations of these 
three variants of spins marks three regimes of dynamics in our cartoon,
as described below.
\\

\noindent
{\bf The Strongly Localized Regime}: This is a regime where a large fraction of the auto-correlation is frozen (Fig.~\ref{Fig5B:Physical_Picture}(a)). This corresponds to the topmost blue (almost frozen) curves in Fig.~\ref{Fig1:Kohlrausch}. In this regime, within $T_{obs} = 10^4,$ the relaxation happens only due to residual small fraction $(1-a)$ of fast (red spins) and slowly decaying (green spins) parts of the auto-correlation 
This corresponds, for example, to the large $\alpha$ regime on the right of the peak of $\tau$ in Fig.~\ref{Fig2:Phenomenology}(b). Here the $L-$dependence of the stretching exponent does not exhibit any systematic trend within system-sizes accessible to us, and stability of the Kohlrausch law cannot be inferred. \\

\noindent
{\bf The Kohlrausch (Slow) Regime and the Occurrence of the Peak in $\tau$:} 
Fig.~\ref{Fig5B:Physical_Picture}(b) shows the situation as we tune the system more towards the delocalized regime either by increasing $\epsilon$ or decreasing $\alpha$ and/or $W.$ A considerable number of frozen (blue) spins are liberated marginally from the strong constraints (that had kept them frozen) to have restricted dynamics and thereby got converted into slow (green) spins. In comparison, the growth of the fraction of the fast (red) spins (presumably, mostly from the already existing slow spins) is not appreciable yet. The dynamics is hence dominated by the slow spins. This corresponds to the curves with intermediate relaxation rates (the pale orange plots in the midway of Fig.~\ref{Fig1:Kohlrausch}), and the region around the peak in $\tau$ in Fig.~\ref{Fig2:Phenomenology}(b). Here the growth rate of the red spins takes over the growth rate of the green spins as one moves towards a more delocalized regime, and  
the peak occurs roughly at the point where the number of red spins overtakes the number of green (and rarely occurring blue) spins. This is the regime (marked by vertical yellow stripe in Fig.~\ref{Fig2:Phenomenology}) where the Kohlrausch law is stable to increase in $L$ -(i.e. $\gamma$ drifts off from unity as $L$ is increased). The stability is concurrent with the dominance of the slow spins.
\\

%

\noindent
{\bf The Delocalized Regime}: Here most of the spins are free to relax fast, resulting in dynamics which is very close to thermal 
(Fig.~\ref{Fig5B:Physical_Picture}(c)). This corresponds to the most rapidly decaying red curves in Fig.~\ref{Fig1:Kohlrausch} and the low $\alpha$ regime on the left of the $\tau$ peak in Fig.~\ref{Fig2:Phenomenology} (b). In this regime the stretching exponent $\gamma$ systematically increases with $L$ approaching unity, indicating an exponential
relaxation for $L\to\infty.$ \\

\noindent
{\bf Numerical Signatures of the Regimes:}  
The frames in the lower panel of Fig.~\ref{Fig5B:Physical_Picture} show the number distribution of spins over auto-correlation (i.e., single-spin  auto-correlation versus the fraction of spins having that particular value of the auto-correlation), over an ensemble of $5000$ spin taken from 5000 various disorder realizations, for each of which
a random $x-$bit-string state (a simultaneous eigenstates of all $\sigma_{i}s$) is taken as the initial state. Each frame in the lower panel corresponds to the physical the situation depicted in the cartoon frame right above it in the upper panel.
The plots illustrate the corresponding cartoons. Fig.~\ref{Fig5B:Physical_Picture} (d) represents the most strongly localized regime,
where there is no significant change in the mean of the distribution (vertical lines representing the ensemble average of the auto-correlation  with colors corresponding to various times) with time, especially in the later decades. Fig.~\ref{Fig5B:Physical_Picture}(e) represents the Kohlrausch regime, where the change in the mean takes place through the later decades, but still a major fraction of the auto-correlation remains intact. Fig.~\ref{Fig5B:Physical_Picture}(f) represents the fast relaxing regime, where rapid decay is observed throughout the entire duration, and also 
shifting the mean from 1 to 0. The consistency of the above behavior with the numerical results for the Kohlrausch phenomenology can be gleaned, at once by recalling, for example, the correspondences of (d), (e), and (f)
with the three regimes -- the one to the right of the yellow band (strongly localized), the one within the yellow band (the Kohlrausch), and the
one to the left of the yellow band (delocalized)
of frames (a), (b), and (c) of Fig.~\ref{Fig2:Phenomenology}.
\\

\subsection{Single-Spin Analytics: A Non-Hermitian Random Hamiltonian Approach}
\label{subsec:Non_Hermitian}
 Here we resort to a simple formulation suitable for capturing 
 the key elements behind the Kohlrausch 
 law and its associated phenomenology in an MBD system.
Since the single-spin auto-correlation is the central object of interest here, the key idea is to focus on the dynamics of a single spin (an arbitrary target spin), treating the rest of the system as a bath. The effects of the disorder, interaction, and drive are thus encapsulated in the action of the bath on the target spin. Here  we resort to a minimal description for the above setting, consisting of a spin evolving under a single-spin non-hermitian Hamiltonian. 
The non-hermiticity is essential for modeling aspects like 
dissipation, energy gain, and dephasing of the dynamics of the target spin.
The stroboscopic dynamics under the time-periodic Hamiltonian can be viewed as stroboscopic observation of the dynamics under the time-independent Floquet Hamiltonian $H_{eff}$ (where the evolution operator over a period $U(T;0) = e^{-iH_{eff}}$), the non-hermitian Hamiltonian, in principle, is to be obtained by tracing over all the degrees of freedom except the target spin, evolving under $H_{eff}$ (time-independent). Thus for stroboscopic dynamics, it is sufficient to consider a time-independent non-hermitian
Hamiltonian whose parameters encode the information of the rest of the system as well as the drive. 
The problem is soluble, and we treat the parameters in the solution as free ones to be extracted from the numerical results.
To this end, we devise a simple single-spin non-hermitian Hamiltonian formulation as follows.\\

\subsubsection{The Formulation}
We describe the dynamics of an arbitrary target spin in our system (observed stroboscopically at $t = nT$) by that governed by a time-independent non-hermitian Hamiltonian as follows.
\begin{eqnarray}
    \tilde{H} &=& H_{h} + H_{nh}, ~ {\rm where} \nonumber \\
    H_{h} &=&  -\Omega\sigma^{x} ~  {\rm and} \nonumber \\
     H_{nh} &=&  -i\left[a_{0} \mathcal{I}_{2\times 2} +  a_{1} \sigma^{x} 
     +  a_{2} \sigma^{y} +  a_{3} \sigma^{z} \right] 
    \label{Eq:H_NonHerm}
\end{eqnarray}
\noindent Here $\sigma^{x/y/z}$ are the Pauli matrices representing the components of the target spin, $a_{i}$  $(i = 1,4)$ are random parameters related to the
parameters of the corresponding random realization
of the time-dependent many-body system in which the
target spin is embedded. Note that when focused
on a single spin, $\Omega$ is just an overall scale factor. However, for an ensemble of spins embedded in
a random many-body system, $\Omega$ will be an 
overall relative scale of relaxations among the spins,
e.g., a larger $\Omega$ will correspond to a faster spin, while a smaller $\Omega$ will correspond to a slower spin, given all other parameters are same between them.
Our approximation
consists of the assumption that $\tilde{H}$
encapsulates the interaction of the target spins with the environment as well as the drive. This is 
a rather drastic approximation, and is suitable 
only for calculating single-spin observables. 
Since we are calculating auto-correlations in $X$-components of the lattice spins, we represent the $x-$component $X_{i}$ of a target spin (say, at a site $i;$ see Eq.~\ref{Eq:FlqHam}) to be diagonal in the $\sigma$ representation, i.e., $(-1)^{n+1}X_{i}(nT) \leftrightarrow \sigma^{z}(nT).$ Note that, since we are representing the dynamics of a spin 
by a time-independent $\tilde{H},$ it is necessary that we incorporate the factor $(-1)^{n+1}$ in the correlations to set the equivalence correctly. \\

We switch to a re-defined set of parameters (after ~\cite{Sergi_2Lvl_Non_Hermitian}) to link them more clearly to various physically intuitive (as has been discussed down the line) quantities as follows.
\begin{equation}
a_{0} = \gamma, ~ 
a_{1} = \gamma \beta, ~ 
a_{2} = \nu, ~ a_{3} = V.
    \label{Eq:Redifined_Param}
\end{equation}
\noindent 
In general, the state of the target spin is 
expected to be represented by a density $2\times 2$ matrix $\rho(t)$ which
satisfies the following Schr\"{o}dinger equation.
\begin{equation}
i\partial_{t}\rho(t) = -i[\tilde{H}, \rho(t)].
    \label{Eq:2_Lvl_TDSE}
\end{equation}
\noindent
Since $\tilde{H}$ is non-hermitian, $Tr[\rho(t)]$ is
not necessarily conserved (here $Tr[.]$ denotes trace). The expectation value of any observable ${\cal O}$ is hence needed to be normalized as follows.
\begin{equation}
\langle {\cal O} \rangle (t) 
= Tr[\rho(t){\cal O}]/Tr[\rho(t)].
    \label{Eq:rho_Normalization}
\end{equation}
The Schr\"{o}dinger equation Eq.~(\ref{Eq:2_Lvl_TDSE}) is exactly soluble~\cite{Sergi_2Lvl_Non_Hermitian} for any real set of values of the parameters in $\tilde{H}$. The general solution reads:\\
\begin{equation}
\begin{split}
\rho_{mn} &= \frac{e^{-\Gamma t}}{2({\beta}^{2} + {\gamma}^{2})}\left[A_{mn}\cos{(\omega t)} + B_{mn}\sin{(\omega t)} \right.\\
 &\left. + C_{mn}\cosh{(\Gamma t)} + D_{mn}\sinh{(\Gamma t)}\right], 
\end{split}
    \label{Eq:Gen_Soln_2Lvl}
\end{equation}
\noindent
where
\begin{equation}
\label{Eq:Gamma_Omega}
 \Gamma = 2\Omega\gamma, ~ \omega = 2\Omega\beta.
\end{equation}
\noindent Here 
$\rho_{mn},$ ($m,n \in \{1,2\}$) are the four elements of the density matrix $\rho(t),$ and $A_{mn}, B_{mn}, C_{mn}$ and $D_{mn}$ are
constants depending on the initial condition and the Hamiltonian parameters for a given disorder realization. This solution is general enough to cover various scenarios including (but not limited to) exponential and polynomial relaxations with or without oscillations, energy-conserving dynamics, pure dephasing, etc as special cases depending on the
values of the Hamiltonian parameters even for a given
initial state (see, 
e.g.,~\cite{Sergi_2Lvl_Non_Hermitian}). This versatility of the exact solution makes it suitable for capturing the diversity of single-spin relaxation (as illustrated in 
Fig.~\ref{Fig:5A_SinglSpnDyn}).
Here we 
focus on a fully polarized eigenstate of 
$\sigma^{z}$ ($X_{i}$) with eigenvalue $+1,$
i.e., all spins up in the $x-$direction in the chain in
Eq.~\ref{Eq:FlqHam}. Under this initial condition,
the time-staggered single-spin auto-correlation
\begin{equation}
\tilde{C}^{i}_{auto} 
= (-1)^{n+1}\langle X_{i}(t) \rangle = 
\langle\sigma_{z}(t)\rangle,
\end{equation}
\noindent
where $t=nT.$ \\

The expression for $\tilde{C}^{i}_{auto}$ (expressed in terms of $\sigma^{z}$) for the above-mentioned initial condition reads (using Equation~\ref{Eq:Gen_Soln_2Lvl})
\begin{widetext} 
\begin{equation}
    \langle\sigma^{z}(t)\rangle_{Analytic} = 
     \frac{[(1 + \gamma^{2} -\nu - V^{2})\cos{(\omega t)}
     + (\beta V)\sin{(\omega t)]\sech{(\Gamma t)} 
+ (\nu + \beta^{2} + V^{2} - 1) 
+ \gamma V\tanh{(\Gamma t)}}}{
         [(\nu + \beta^{2} - 1)\cos{(\omega t)} 
+ (\beta V)\sin{(\omega t)}]\sech{(\Gamma t)}  
          + (1 + \gamma^{2} -\nu) + \gamma V\tanh{(\Gamma t)}},
 \label{Eq:Sz_Analytical}
\end{equation}
\end{widetext}
\noindent
The expression contains the parameters $\Omega, \gamma, \beta, \nu$ and $V$, that are essential for
characterizing a $2\times 2$ non-hermitian matrix 
(Eqs.~\ref{Eq:H_NonHerm} and \ref{Eq:Redifined_Param}),
and are expected to encode the necessary effects of the
drive and the other spins on the relaxation of the target spin.\\
\begin{figure*}[ht]
\begin{center}
\includegraphics[width=0.32\linewidth]{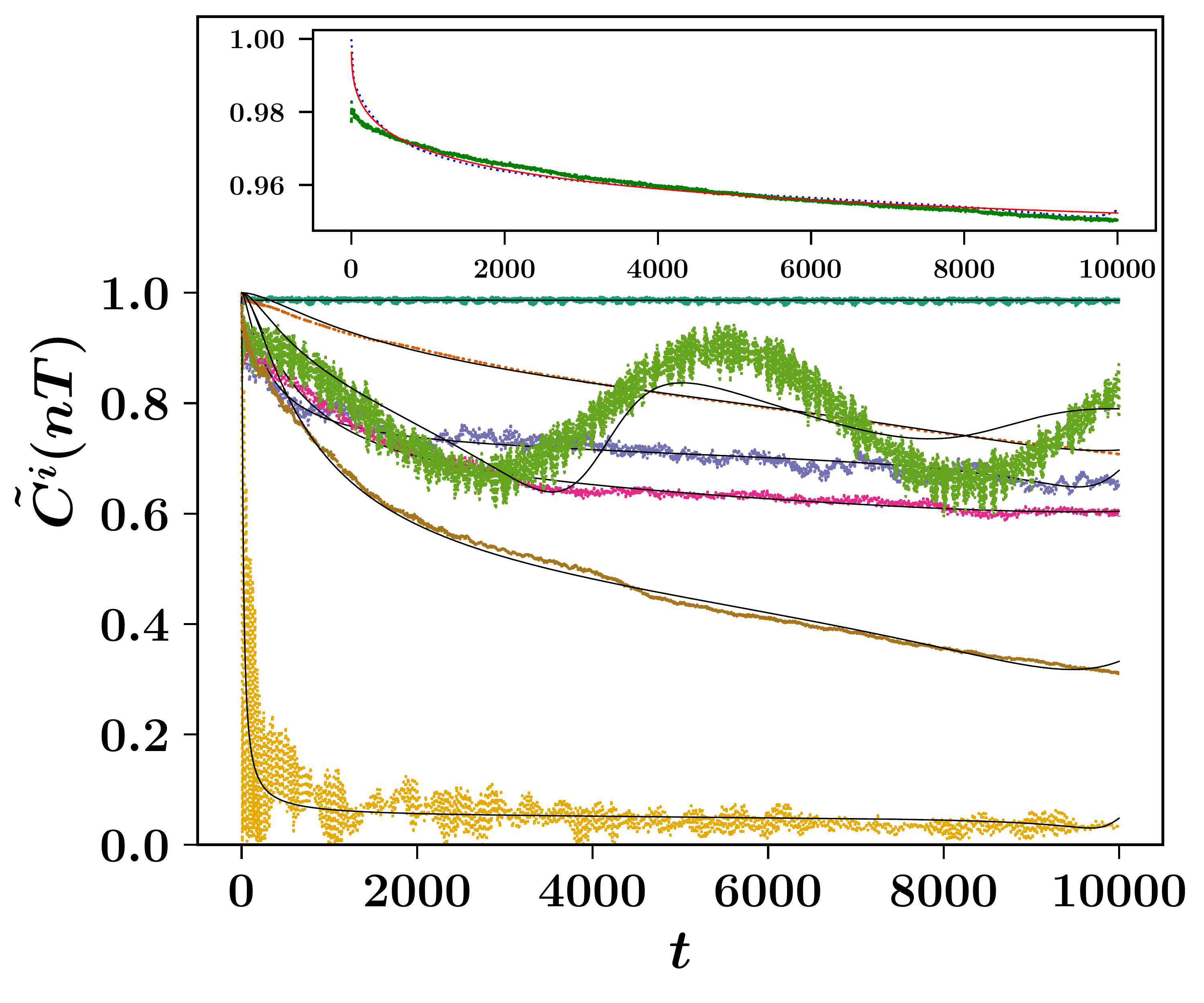}
 \includegraphics[width=0.32\linewidth]{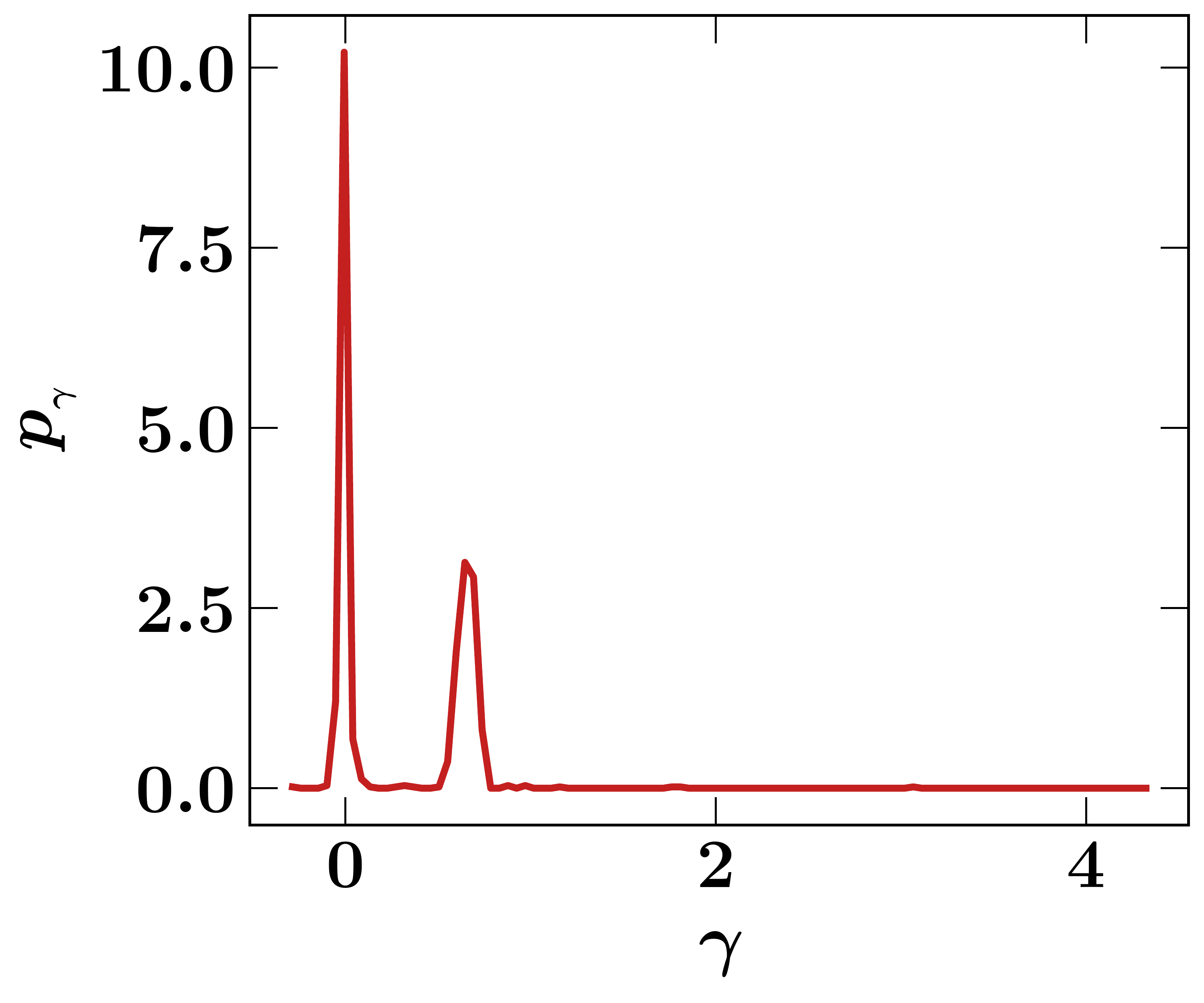}
\includegraphics[width=0.32\linewidth]{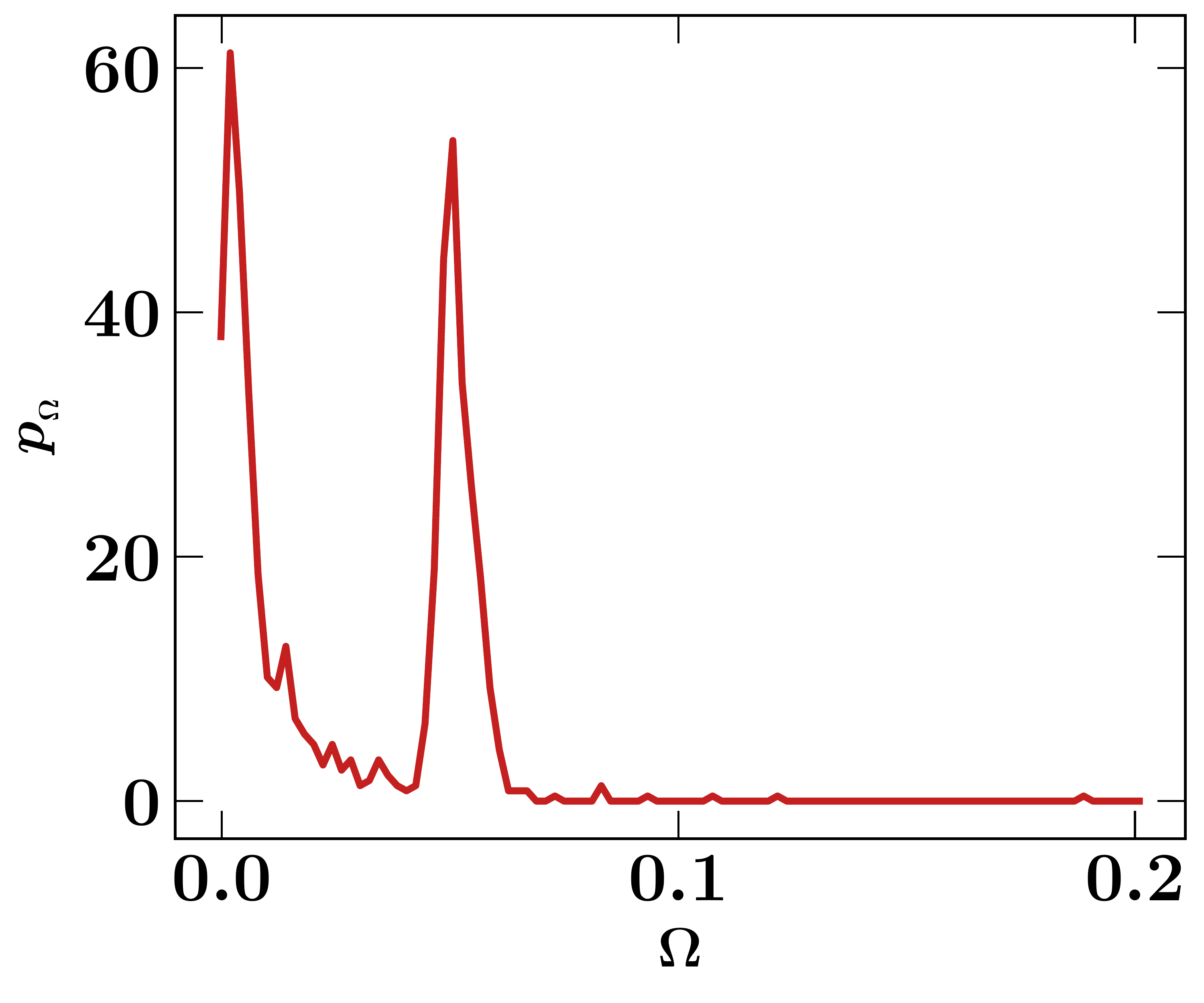}

\end{center}
\caption{Fitting of the exact numerical results for the time staggered
single-spin auto-correlations with the
analytical formula from the non-hermitian formulation.
{\bf Left:} main frame shows some representative fitting of the numerical data (colored) by the analytical function (solid black lines). The inset shows the fitting of the numerical disorder averaged relaxations with the disorder averaged analytical expression corresponding to the
case $\alpha =2.8 , W = 0.4\pi, \epsilon =0.2 , L = 26$ (Kohlrausch regime). The fitting is quite accurate and satisfies Kohlrausch's law. 
{\bf Middle:} frame shows the probability distributions $P_{_\gamma}$ of $\gamma.$
{\bf Right:} frame shows the probability distributions $P_{_\Omega}$ of $\Omega.$ 
}
\label{Fig:Non_Herm_Kohlrausch}
\end{figure*}
It is easy to glean from Eq.~(\ref{Eq:Sz_Analytical}) that $\gamma$ sets the time-scale of energy loss/gain
by the target spin, $\beta$ provides the characteristic
frequency of the damped oscillation executed by the target spin. \\

Since energy is pumped into the system, there
are generically exponentially growing terms in the 
subsystem dynamics, as well as
exponentially decaying terms due to both drive and
the loss of energy to the rest of the system.
A steady state appears when those
two rapidly changing terms counterbalance each other.
In the above form (Eq.~\ref{Eq:Sz_Analytical}), the expression for $\SzAna$ this is in-built, and
 averts the usual problem with divergences associated with individual exponentially increasing terms as $t\to\infty$ and the associated numerical problems at the large but finite time: the magnitude of both $\sech{(x)}$ and $\tanh{(x)}$ are bounded between $0$ and $1$. 
 
\subsubsection{Auto-Correlation Dynamics and the Kohlrausch law from the Analytics}
We fit the single-spin auto-correlation data obtained from exact numerics (as shown in Fig.~\ref{Fig:5A_SinglSpnDyn}) with the analytical expression for $\SzAna$ in Eq.~(\ref{Eq:Sz_Analytical}), for a large number of disorder realizations, and extract the values of the parameters $\Omega, \gamma, \beta, \nu,$ and $V$ in each case. Plugging such a set of parameter values into the expression of $\SzAna,$ we get our analytical expression for $\tilde{C}_{i}(t)$ with numerical values for its parameters. We average this expression over various disorder realizations. \\

The large varieties of single-spin relaxations
obtained numerically (Fig.~\ref{Fig:5A_SinglSpnDyn})
is captured by our analytic expression obtained from a simple formulation (Fig.~\ref{Fig:Non_Herm_Kohlrausch}, {\bf Left} frame, main) for a few random samples. The inset shows the fit of the 
disorder-averaged numerical result by the disordered averaged analytical expression. They are both
consistent with the Kohlrausch law to very good accuracy
over multiple decades.\\

The probability distribution $P_{_{\gamma}}$ of the relaxation rate $\gamma$ ({\bf middle} frame) also shows two clear peaks.
 The population at the peak at $\gamma = 0$ exhibits no relaxation at all (blue spins), while those within the width of it show slow relaxation (green spin), and the
 second peak represents spins with considerable relaxation. The bimodal structure appearing in the 
 distribution of these two parameters with an
 appreciable width around the peak at zero is consistent with the classification
 of our spins into three categories, namely, the frozen (blue), the slow (green) and the fast (red)
 based on their relaxation rates. The distributions
 for $\nu,$ $\beta$ and $V$ are shown in Appendix~\ref{App:Prob_nu_beta_V}.
Similarly, 
the distribution of $\Omega$ (Fig.~\ref{Fig:Non_Herm_Kohlrausch}, {\bf  right} frame) exhibits two clear peaks - one at $\Omega \approx 0$, another around $\Omega \approx 0.06$. The peak at 
$\Omega = 0$ represents the spins frozen to our resolution (the blue spins). The populations within the
width (suitably defined) of this peak represent the slow (green) spins. The second peak and those within its width represent spins with faster relaxations (the red spins in our cartoon). 


\section{Early vs Late Stages of Relaxation: A Numerical Search for the ``critical" DTC Phase}
\label{SecV:NoCDTC}
\begin{figure}[ht]
\begin{center}
\includegraphics[width=0.95\linewidth]{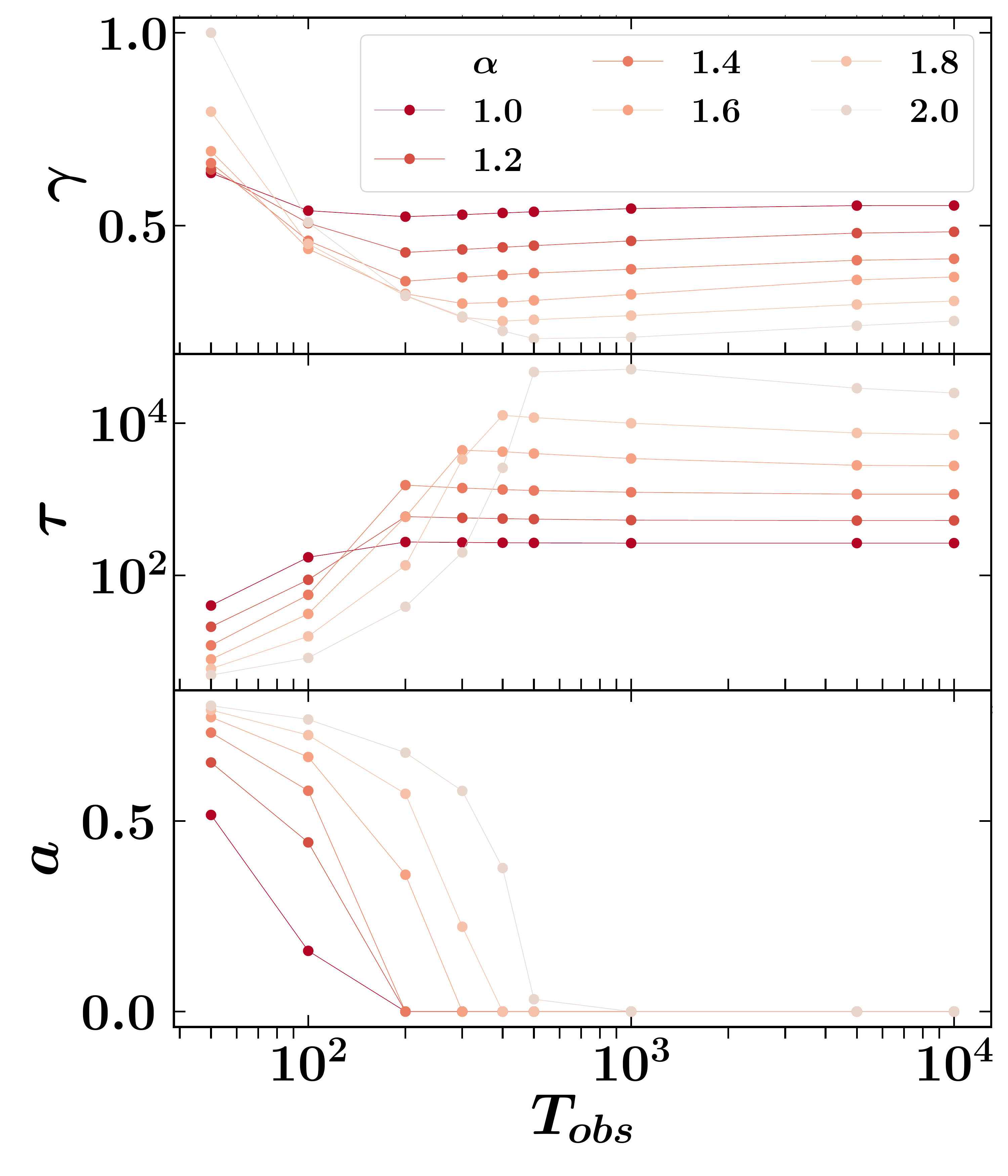}
\end{center}
\caption{
Dependence of the relaxation parameters with the observation time $T_{obs}$ indicates the existence of two distinct dynamical stages. At the early stage (roughly corresponding to $T_{obs} < 500$), the relaxation parameters are quite sensitive to the observation time $T_{obs}$ (especially for the shorter-ranged systems). At the later times (the late stage), the relaxation parameters tend to saturate and their dependence on $T_{obs}$ is much weaker or negligible.
}
\label{Fig:Var_wth_Tobs}
\end{figure}
The existence of various stages of relaxation is not uncommon in interacting many-body systems. For example, a system described by a fractional Fokker-Planck equation can exhibit relaxation that crosses over from stretched exponential to an inverse power law (power law with negative exponent)~\cite{Double_Regime_Stretched_InvPowr}. 
Here we found, quite interestingly, that the stretched exponential form of Eq.~\ref{Eq:Fit_Function} continues to fit both the early and late stages the best (compared to other more common relaxation forms like exponential and power-laws) and also remarkably accurately. The fitting parameters are distinctly different for the two stages distinguising 
one from the other as shown in Fig.~\ref{Fig:Var_wth_Tobs}. 
One should however keep in mind, that the time-scale of
the emergence of the late regime is still negligibly small 
compared to the characteristic time-scale of relaxation (if any) of the frozen (blue) spins (the latter being infinity to our resolution).
\\



The phenomenology of relaxation in the early and late stages is distinct as discussed below. Fig.~\ref{Fig:Var_wth_Tobs} (top frame) shows, initially the stretching exponent $\gamma$ is closer to unity (i.e., the relaxation is more thermal-like) for larger $\alpha$ (shorter interaction range). But between $T_{obs} = 100$ and $200,$ the trend reverses, and $\gamma$ drifts further from unity for larger $\alpha.$ The distinction between the two stages of relaxation is most pronounced for the short-ranged systems and gradually levels off as the range of interactions is increased. This can be roughly understood as follows. 
In shorter-ranged MBD systems, for a sufficiently strong disorder, there is localization (regardless of whether it persists as $t\to\infty$) at the initial stage of the dynamics (assuming we start with a localized state). The early stage consists of dynamics within the associated localization length. This is followed by a much slower relaxation process involving equilibration between different localized regions. With an increase in the range of the interaction, the size of the localized patches increase. The relaxation dynamics hence involve cooperative dynamics of a larger number of  spins. This makes the relaxation slower than the shorter-range system at early times (see, e.g., ~\cite{Lea_Santos_Bar_Lev_Slow_LongRange}). The early stage is thus characterized by faster relaxation (larger $\gamma$) for the shorter-ranged systems. The late stage is marked, on the other hand, by the equilibration dynamics between the localized patches (beyond the localization length), which is much slower in shorter-range systems due to stronger localization 
compared to the longer-ranged ones. This is hence characterized by a smaller value of $\gamma$ (more stretched relaxation) for shorter-ranged systems. \\

The persistence of Kohlrausch law in almost all stages of relaxation is also interesting in view of a prediction of a dynamical phase called ``critical" DTC in long-range interacting Floquet systems with unitaries of the form given in Eq.~\ref{Eq:FlqUnitary} in various spatial dimensions\cite{Dima_CDTC, Else_Norm_DTC_Review}. A perturbative expansion of an effective Hamiltonian in a suitable frame under specific conditions predicts a scale-free relaxation of a DTC order parameter (the auto-correlation $\tilde{C}$ we measured) at the early time~\cite{Dima_CDTC,Else_Norm_DTC_Review} in a power-law interacting setting whose one-dimensional version we studied (Eq.~\ref{Eq:FlqUnitary}). 
The dynamical phase is called critical DTC, and is predicted to occur 
when the exponent $\alpha$ of the power-law interaction equals the dimension of the lattice ($\alpha=1$ for our setting). \\

\begin{figure}[ht]
\begin{center}
\includegraphics[width=0.95\linewidth]{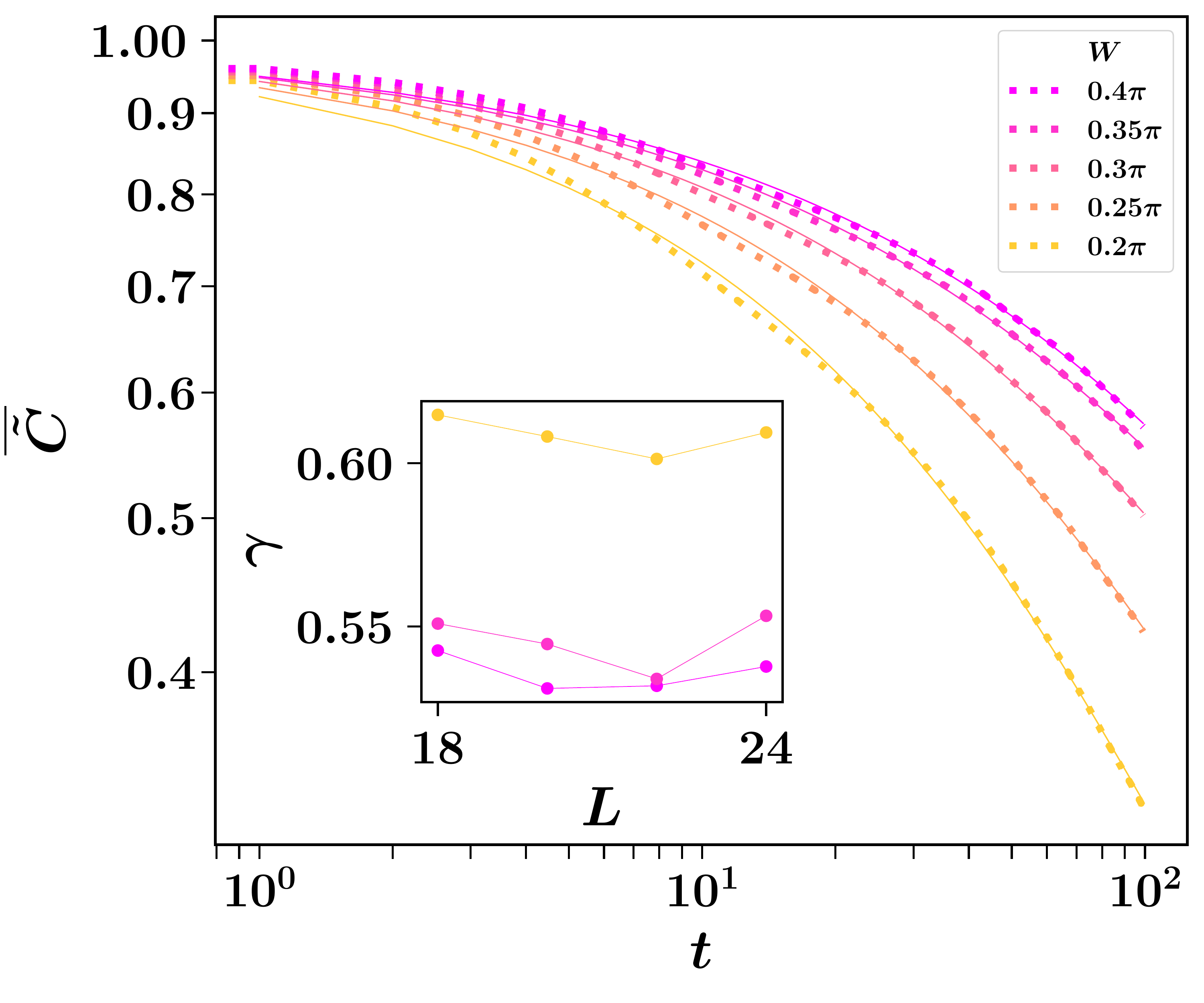}
\includegraphics[width=0.95\linewidth]{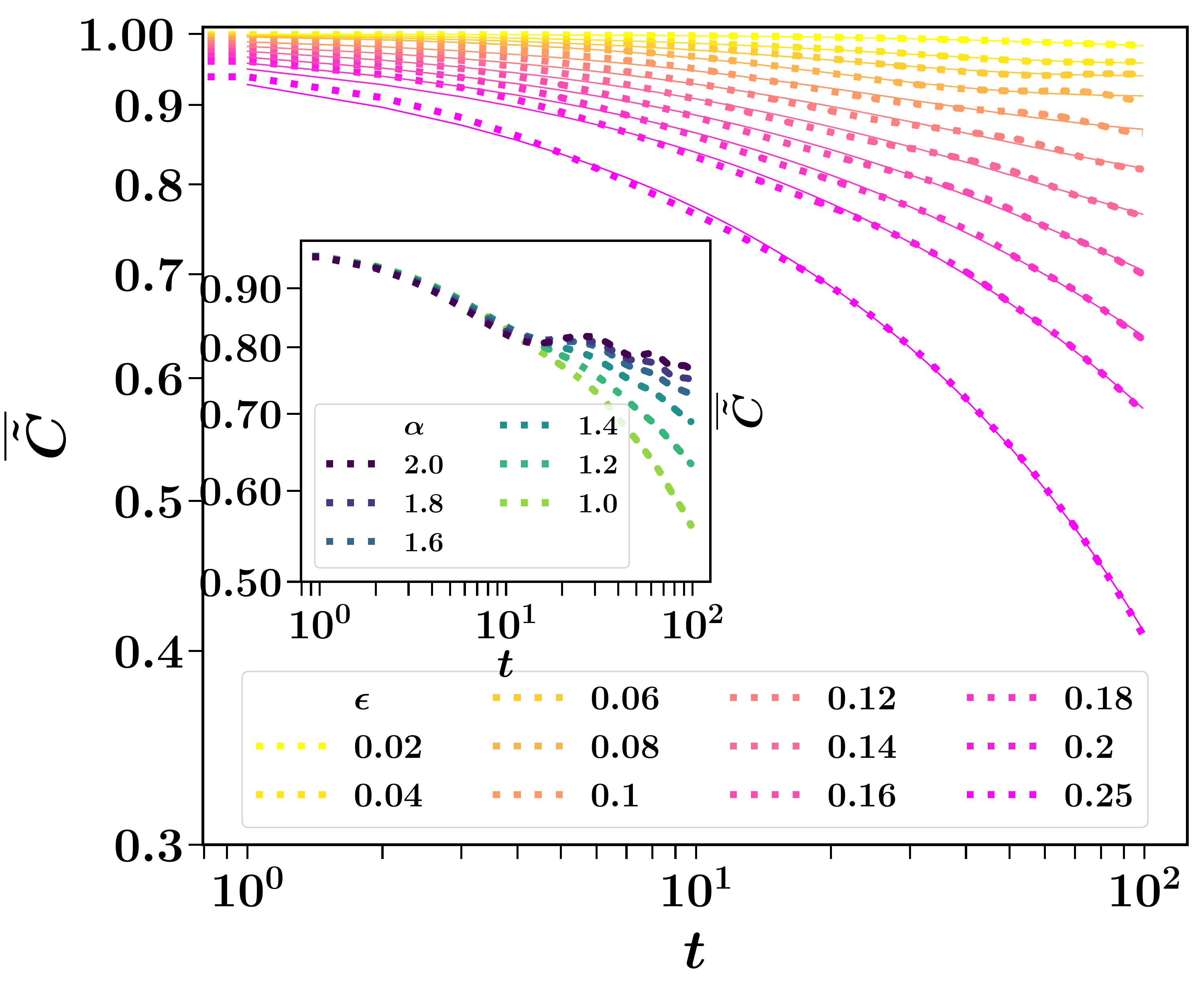}
\end{center}
\caption{
Short-time behaviors and fitting: Showing no sign of power-law decay. A log-log plot does not show a straight line and fits better (least squares being always $< 10^{-6}$) with a stretched exponential function (We tried variants of exponential fits, which always gives least squares $~10^{-2}$, data not shown). \textbf{Top Panel:} Main plot is for $L=24,\epsilon=0.2, \alpha=1.0$ and the inset shows the system size scaling of the exponent $\gamma$ for various values of $W.$  \textbf{Bottom Panel:} Main plot is for $L=24,W=0.4\pi, \alpha=1.0.$ The {\it inset} shows the absence of power-law relaxation for a range of 
values of $\alpha$
around $\alpha=1.$
}
\label{Fig:NoPower}
\end{figure}

Focusing on the early-time data (the first two decades) we could not find the proposed critical DTC phase in spite of a careful scanning of the parameter space. Instead, we found Kohlrausch law persists across the board for $\alpha=1,$ as shown in Fig.~\ref{Fig:NoPower}. \\

\section{Discussions and Outlook}
\label{Outlook}
\noindent
{\bf Spin Glass Mechanism Plays No Essential Role in Kohlrausch Stretching of Relaxation:}
Glass formation is known to one of the key mechanisms that give rise
to stretched relaxation (see, e.g.~\cite{StrchdExp_MolElec_Glasses_Review}). In our spin set-up,
formation of a quantum spin glass (see, e.g., \cite{Quantum_SpinGlass_Rieger_Young} for a review) could hence have been a candidate behind the Kohlrausch relaxation. However, two
essential ingredients of a spin glass are (a) disorder and (b) frustration~\cite{Parisi_SpinGlass_Book,Sei-Book}. Our results indicate that Kohlrausch stretching of relaxation is in fact independent of the latter. \\

In our setting, frustration is introduced if a fraction of the interactions $J_{ij}$ are made negative in the long-range case -- there is no frustration when all the bonds are ferromagnetic, while the presence of a sufficient fraction of anti-ferromagnetic bonds can introduce the frustration necessary to stabilize a spin-glass in a long-range interacting disorder system~\cite{Binder_Young_RMP}. We find that
Kohlrausch stretching is present even in systems with completely ferromagnetic couplings ($J_{ij} \ge 0$). As a strong example, 
we analytically show (see Appendix~\ref{app:frustration}) that taking any arbitrary bit-string state in $x$ direction as the initial state, 
the dynamics with all $J_{ij} > 0,$
in the interaction part $H_{x}$ (Eq.~\ref{Eq:FlqHam}), i.e., completely 
ferromagnetic interactions are exactly identical for that in which each $J_{ij}$ has flipped sign (highly frustrated one).
\\

\noindent
{\bf On Plausible Saturation of the Peak with increasing $L$:} 
The fraction of frozen spins observed within $T_{obs}$ depends on $L.$ All we assume is that it asymptotes to some non-zero value even as $L\to\infty.$ This only requires the characteristic time for melting
of the Local Integrals Of Motion/$\ell-$bits (see, e.g., ~\cite{Abanin_MBL_Review}) to be larger than $T_{obs}.$ 
Though we have not seen the saturation within the system size we can access numerically, on physical grounds, the assumption seems quite justified for strong disorder - the thermalization is shown to be extremely slow even where MBL is argued to be absent~\cite{Anatoli_Sels_NoMBL}, and hence the frozen part of the correlation is expected to persist even for substantially long $T_{obs}$. \\

As $L$ is increased, the peak in $\tau$ grows in
height (as can be seen from Fig.~\ref{Fig2:Phenomenology}(b)). 
However, for a finite $T_{obs},$ the peak height will eventually saturate beyond some very large value  $L \sim L_{max}$ (not accessible to numerics). This may be understood as follows. 
We assume that the number density of spins with effective relaxation time-scale $\tau$ is $n(\tau),$ than the average separation between such spins will be $1/n(\tau).$ Then it is easy to see that the average height of the peak can grow as $L$ is increased till some 
$L_{max} \sim 1/n(\tau_{max})$ where
$\tau_{max}$ is the  smallest value of $\tau$ for which $\int_{\tau_{max}}^{\infty} \tau n(\tau) d\tau \approx 0$ (provided such a $\tau_{max}$ exists).
Till $L$ reaches $L_{max},$ increasing $L$ might 
result in an inclusion of much slower spins (much larger $\tau$) with an appreciable
 probability. This might lift the average value of $\tau$ and hence the peak height 
 (as it does in our case). 
In the following, we argue the existence of
a $\tau_{max}$ in our setting.
Since we assumed a finite $T_{obs}$ and finite
resolutions of the auto-correlation measurement,
we cannot resolve a very large $\tau$ from infinity,
from the auto-correlation measurement (for example, for $T_{obs} = 10^{4},$ a spin with $\tau = 10^{15}$ will be considered as frozen with a resolution that cannot register any change in correlation taking place over a period of $10^{-11}\times\tau$). The cut-off value of $\tau$ beyond which the change in auto-correlation
cannot be registered thus serves as a natural
$\tau_{max}.$ We consider spins with $\tau > \tau_{max}$ as completely frozen and hence 
non-participating 
in the measurable
correlation decay (their contribution constitutes the atemporal part $a$ of the auto-correlation).
Hence, taking into account only the remaining spins,
we have $n(\tau) = 0$ for $\tau > \tau_{max}.$
Since this is indeed the case ($a \ne 0$) in the
Kohlrausch regime where the peak occurs, a $\tau_{max}$ exists in this regime for which
$\int_{\tau_{max}}^{\infty}\tau n(\tau)d\tau =0,$
and the peak will saturate beyond $L_{max}$ associated to it. \\


 
 The peak also moves deeper into the more
 localized regime as $L$ is increased. That could simply happen because the putative MBL phase seems to move further away with increasing system size~\cite{Prosen_Suntaj_NoMBL_1,Anatoli_Sels_NoMBL,Morningstar_NoMBL} in most cases. However, since we are focusing on a finite $T_{obs},$ by our assumption that there will always be a residual auto-correlation even in the thermodynamic limit for $t < T_{obs},$ the peak should converge to some finite value as $L\to \infty.$ Unfortunately, this saturation is also beyond the system sizes accessible to our numerics. \\
 
 \noindent
 {\bf The Peak as $T_{obs} \to \infty:$} If there is a stable MBL for a finite value of the disorder strength, then the above picture in principle can hold even as $T_{obs}\to\infty.$ This is because  there will always be some blue spins, and it is plausible that as one moves towards a less localized regime from a more localized one, some of them will be liberated to give rise to the green spins. Of course, there will be a competition between the infinities: the infinity associated with the time-scale of stability of the MBL and the infinity associated with $T_{obs}.$
 On the other hand, if there is no stable MBL anywhere, then as $T_{obs}$ is increased further, more blue spins will actually qualify as the green spins, and the peak height will increase,
 and eventually saturates to the longest delocalization time-scale as $T_{obs} \to \infty$. \\

\noindent    
\begin{center}
    {\bf Conclusions:}
\end{center}

    \noindent
    We showed that the century-old law for relaxation of complex classical systems (structural glass, super-cooled liquid, etc) -- the famous Kohlrausch law -- also appears ubiquitously in the domain of isolated MBD systems -- both driven and time-independent -- evolving under pure 
    Schr\"{o}dinger dynamics. \\

     \noindent
     This helps us quantify the slowness of the well-known prethermal
     relaxation in MBD systems. \\
          
    \noindent
    Finite-size analysis shows systematic stabilization of this law with increasing system size in certain parameter regimes, allowing us to define a
    dynamical phase characterizing the effect of disorder. \\
    
    \noindent
    The relaxation time $\tau$ of auto-correlation over a finite observation time $T_{obs}$ shows a peak between the weakly and most strongly localized (characterized by the frozen part of the auto-correlation as appears from a finite time observation) regimes.  \\
    
    \noindent
    The phenomenon does not necessarily depend on the existence
    of a stable MBL phase in the thermodynamic limit at a finite parameter strength
    (e.g., disorder strength $W$, power $\alpha$ of the power-law interaction, etc) that controls the degree of localization, though, the existence is assured if
    there is a stable MBL phase. \\

    \noindent
    We provide a simple analytical formulation
    of the problem, targeting a single spin in the MBD system. The approach is based on random non-hermitian Hamiltonians. Our analytical formula provides an accurate description of the single spin dynamics even at the level of each individual target spin and consequently captures the Kohlrausch law very accurately upon disorder averaging. The characteristics of the MBD phase are captured through the behavior of the system
    as a bath to a random target spin, and
    are encapsulated in the distribution of the parameters in the single-spin non-hermitian Hamiltonian. \\


\noindent
\begin{center}
    {\bf Outlook:} 
\end{center}

    \noindent
    The existence of some sort of slow dynamical phase beyond the MBL regime is probably already 
    in our indirect knowledge: It was known that one sees
    logarithmic growth of entanglement in the MBL phase~\cite{Log_Entanglement}, but now
    it is known that MBL is unstable in a
    substantial part of that regime~\cite{Prosen_Suntaj_NoMBL_1,Prosen_Suntaj_NoMBL_2,Morningstar_NoMBL}. Our
    results are hence expected to trigger 
    a new direction of study toward understanding disordered material in terms
    of new dynamical phases. \\
    
    \noindent
    The diversity of single spin relaxations, especially in the moderately localized regime where the Kohlrausch law is stable, 
    indicates that the many-body bath appears to exhibit a broad range of behaviors to various spins -- ranging from sustained oscillatory to over-damped. Determining the distribution of such bath characteristics might be an interesting way to characterize an MBD system. \\
    
    \noindent
    Investigating the Markovianity of the dynamics and information scrambling in the Kohlrausch phase might reveal interesting facets of 
    relaxation in MBD systems. \\
    

\vspace{0.75cm}

\noindent
{\bf Acknowledgments:} The author is deeply grateful to 
V. Khemani and R. Moessner for several crucial discussions at various stages of the project starting
right from its inception, as well as for their continuous encouragement and support. The author also thanks A. Chandran, A. Das, A. Polkovnikov and T. Prosen for many useful discussions.

\appendix

\section{The Probability Distributions of $\beta,$ $\nu$ and $V$}
\label{App:Prob_nu_beta_V}
\begin{figure}[ht]
\begin{center}
\includegraphics[width=0.523\linewidth]{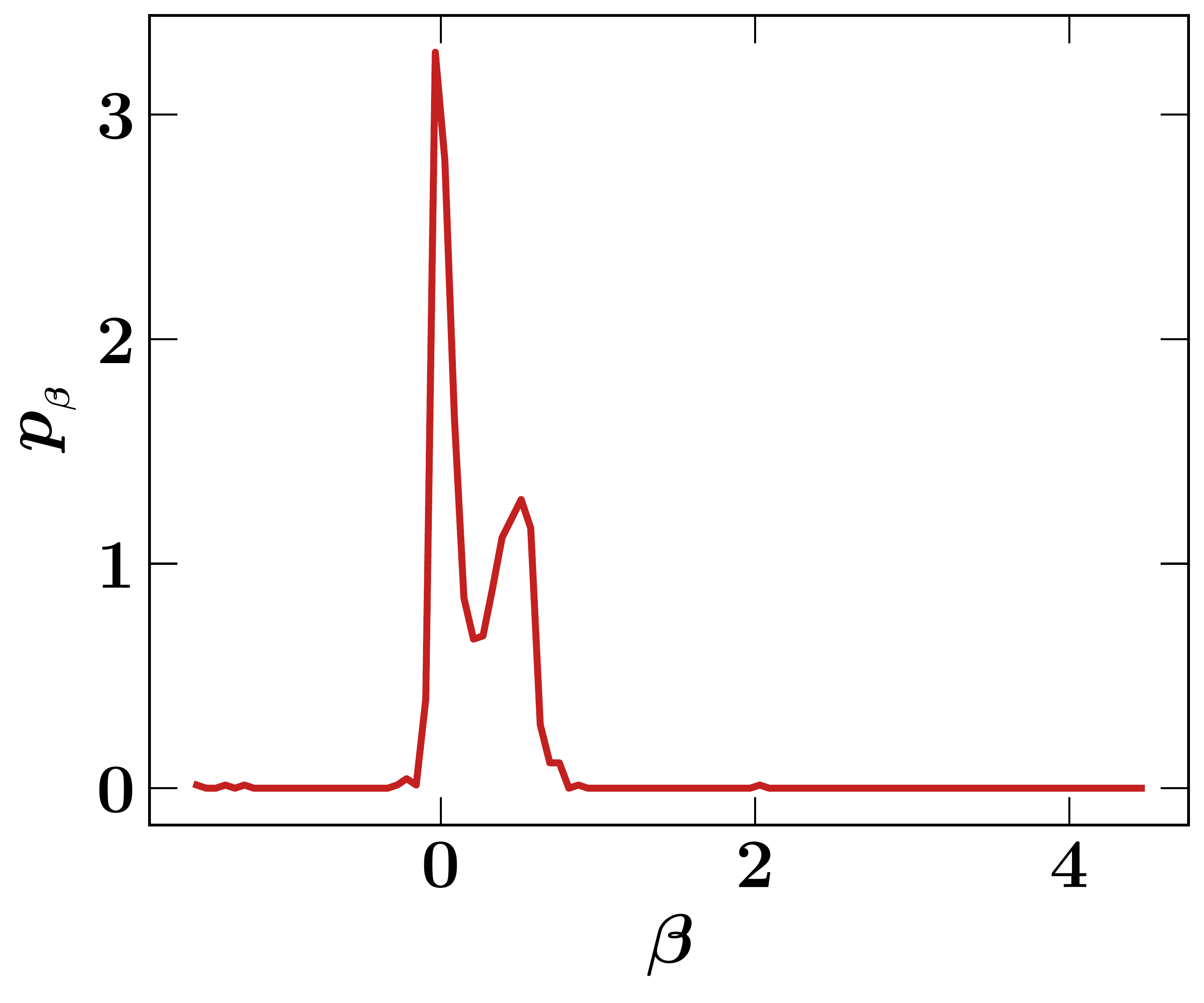} 
   \includegraphics[width=0.523\linewidth]{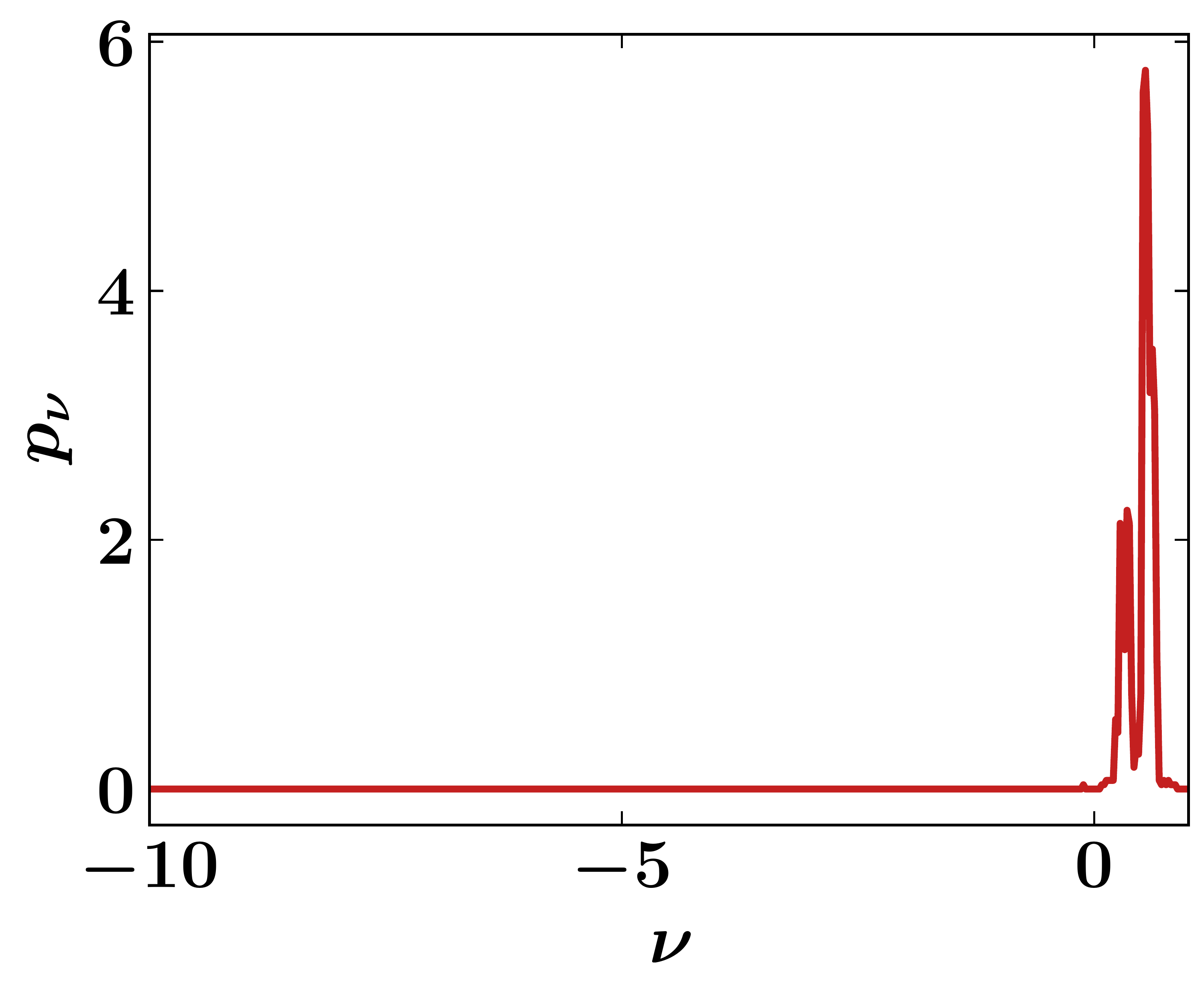}
   \includegraphics[width=0.523\linewidth]{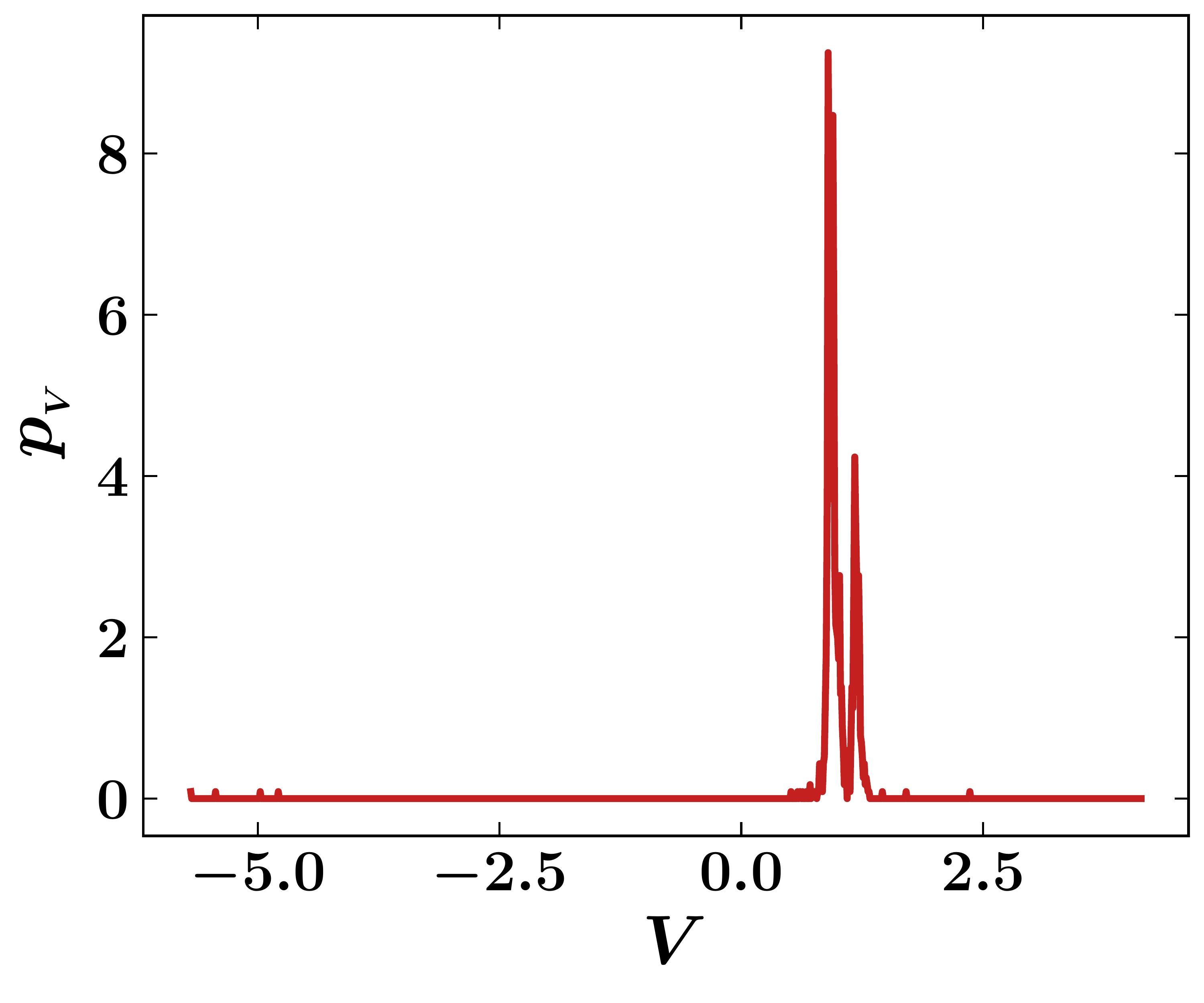}
   \end{center}
\caption{
The Probability Distribution for $\beta,$ $\nu$
and $V$ corresponding to the
case $\alpha =2.8 , W = 0.4\pi, \epsilon =0.2 , L = 26$ (Kohlrausch regime). The results are obtained (for this Fig. as well as the later ones) by averaging over 5000 disorder realizations for each data point, and for each realization, a random $x$-bit-string state is taken as 
the initial state.
}
\label{Fig:App_Prob_beta_nu_V}
\end{figure}
Here we provide the probability distributions of 
some of the parameters of the non-hermitian Hamiltonians
obtained from the fitting of the analytical formula
(Eq.~\ref{Eq:Sz_Analytical}) to the numerical results
obtained for various disorder realizations.

\section{Ruling out Spin-Glass Mechanism as Essential in Kohlrausch Stretching:}
\label{app:frustration}
In order to strike out any role of frustration in the slow dynamics in our setting, we show that taking any arbitrary bit-string state in $x$ direction as the initial state, 
the dynamics with all $J_{ij} > 0,$
in the interaction part $H_{x}$ (Eq.~\ref{Eq:FlqHam}), i.e., completely 
ferromagnetic interactions are exactly identical for that in which each $J_{ij}$ has flipped sign (highly frustrated one). Our proof subsumes exact dynamical equivalence between a system with random ferromagnetic interactions and that with 
Sherrington-Kirkpatrick (SK) spin glass interactions~\cite{SK_Model} in the above
sense for each of the ($2^L$) eigenstates of the $\{\sigma^{x}_{i=1,L}\}$ as the initial states
having all kinds of energies. For a glassy system, a low energy initial state is generically expected to get trapped in extremely long-lived metastable states (surrounded by lofty energy barriers) during relaxation, while in a random ferromagnet, the dynamics do not slow down due to such long-lived glassy traps. Here we show that the slowing down happens equally
in the case of a glassy $H_{x}$ (being completely antiferromagnetic, long-ranged, and random~\cite{Inoue_BKC}) as well as a purely random ferromagnetic $H_{x},$ ruling out any essential role of the frustration in $H_{x},$ in the stretching of the relaxation.
\\

To show this, 
we consider Floquet evolution of an observable $O_{x}$ diagonal on $x$-basis under a unitary (over a period) of the form
  \bea
  U &=& U_{z}U_{x}, ~ {\rm where} \nonumber \\
  U_{x} &=& e^{-iH_{x}(\{\six\})}, ~ U_{z} = e^{-i(\pi + \epsilon)\sum_{j}\sigma_{j}^{z}},
  \label{U}
  \eea  
  \noi and show that its dynamics exactly maps to that of $O_{x}$ evolved with
  \bea 
  \tilde{U} &=& U_{z}\tilde{U_{x}}, ~ {\rm where} \nonumber \\
  \tilde{U}_{x} &=& e^{+iH_{x}}
  \label{U_tilde}
  \eea  
  \noi (that is, with a sign-flipped $H_{x}$), provided the initial states $|\psi(0)\rangle$ (for evolution with $U$) and $|\tilde{\psi}(0)\rangle$ (for evolution with $\tilde{U}$) satisfy the following condition:
   \beq 
  [a_{\al}^{(0)}]^{*} = e^{-i\ga_{\al}^{(0)}}\tilde{a}_{\al}^{(0)},
  \label{Init_Cond}
  \eeq 
  \noi for some $\ga_{\al}^{(0)},$ where
  \bea 
  |\psi(0)\rangle &=& \sum_{\al}a_{\al}^{(0)}|x_{\al}\ra ~ {\rm and} \non \\
  |\tilde{\psi}(0)\rangle &=& \sum_{\al}\tilde{a}_{\al}^{(0)}|x_{\al}\ra, 
  \label{psi_0}
  \eea  
  \noi 
  Here $H_{x}$ denotes the interaction part that depends only on $\six$s, and
  $|x_{\al}\ra$ denotes the $x$-basis states. {\bf Note} that this condition includes the same initial states (e.g., same random bit-string $\xal$)
  for the dynamics under $U$ and $\tilde{U}$.  \\
 Taking $H_{x} = -\sum_{i>j}J_{ij}/r_{ij}^{\al}\six\sjx$ with $J_{ij} > 0$ 
 \\

  We prove this by induction. We assume the states after the $n-$th complete 
  drive cycles are 
  \bea 
  |\psi(n)\rangle &=& \sum_{\al}a_{\al}^{(n)}|x_{\al}\ra ~ {\rm and} \non \\
  |\tilde{\psi}(n)\rangle &=& \sum_{\al}\tilde{a}_{\al}^{(n)}|x_{\al}\ra, 
  \label{psi_n}
  \eea 
  \noi with the coefficients satisfying
  \beq 
  [a_{\al}^{(n)}]^{*} = e^{-i\ga_{\al}^{(n)}}\tilde{a}_{\al}^{(n)},
  \label{nth_Cond}
  \eeq 
  then one can show (see~ appendix\ref{app:frustration} ) that after the $(n+1)-$th cycle, the coefficients (for the expansion of $|\psi(n+1)\rangle$ and $|\tilde{\psi}(n+1)\rangle$) will satisfy
  \beq 
  [a_{\al}^{(n+1)}]^{*} = e^{-i\ga_{\al}^{(n+1)}}\tilde{a}_{\al}^{(n+1)},
  \label{n+1_th_Cond}
  \eeq 
  \noi {\bf Note} that Eq.~\ref{n+1_th_Cond} automatically implies equality of the mod-squares of the coefficients of the individual $x-$basis states between the states evolved by $U$ and $\tilde{U}$ respectively at all time, hence the equality of $\la O_x \ra (n)$ in both cases. \\

 We denote
  \bea
  U_{x}\xal &=& e^{-i\phi_{\al}}\xal ~ {\rm and} \\
  U_{z}\xal &=& \sum_{\be} \balb\xbet. 
  \label{Defs_balb}
  \eea 
  \noi Then we have
  \bea 
  |\psi(n)\rangle &=& \sum_{\al}a_{\al}^{(n)}|x_{\al}\ra ~ {\rm and} \non \\
  |\tilde{\psi}(n)\rangle &=& \sum_{\al}\tilde{a}_{\al}^{(n)}|x_{\al}\ra \non \\
  &=& \sum_{\al}e^{i\ga_{\al}^{(n)}}[\tilde{a}_{\al}^{(n)}]^{*}\xal. 
  \label{psi_n_2}
  \eea 
  \noi Last step is obtained using Eq.~(\ref{nth_Cond}). Now applying
  $U$ on the expression of $|\psi(n)\rangle$ above, we get 
  \bea
  |\psi(n+1)\ra &=& \sum_{beta} a^{(n+1)}_{\be}\xbet, ~ {\rm where} \\
  a^{(n+1)}_{\be} &=& \sum_{\al} \aal^{(n)}e^{-i\phi_{\al}}\balb
  \label{U_psi_n}
  \eea 
  \noi Similarly, applying $\tilde{U}$ on $|\tilde{\psi}(n)\rangle$ in Eq.~\ref{psi_n_2}
  \bea
  |\tilde{\psi}(n+1)\ra &=& \sum_{beta} \tilde{a}^{(n+1)}_{\be}\xbet, ~ {\rm where} \\
  \tilde{a}^{(n+1)}_{\be} &=& \sum_{\al} [\aal^{(n)}]^{*}e^{i\phi_{\al}}e^{i\ga_{\al}^{(n)}}\balb.
  \label{Utilde_psi_n}
  \eea 
  
  \noi 
  Now the strategy is to put the above expressions in the expressions
  (Eq.~\ref{n+1_th_Cond}) we want to prove and see if we get something correct (that is, an assignment of $\gamma_{\al}$'s that satisfies it) or something absurd (we find no such assignment possible).
  Eq.~(\ref{n+1_th_Cond}) to be true, we must have (using Eq.~\ref{Utilde_psi_n} in it),
  \bea
  [a_{\be}^{(n+1)}]^{*} 
  = e^{-i\ga_{\be}^{(n+1)}}\sum_{\al} [a_{\al}^{(n)}]^{*}e*{i\phi_{\al}} e^{i\ga_{\al}^{(n)}}\balb.
  \label{3C}
  \eea 
  Comparing this with the complex conjugate of Eq.~\ref{U_psi_n}, we must have
  \beq 
  \sum_{\al} [\aal^{n}]^{*}e^{i\phi_{\al}}[\balb]^{*} 
  = e^{-i\gamma_{\be}^{(n+1)}}\sum_{\al}e^{-i\gamma_{\al}^{(n)}}[\aal^{n}]^{*}e^{i\phi_{\al}}\balb.
  \label{Sum}
  \eeq 
  Since $\phi_{\al}$ should depend on the specifics of $H_{x},$ in order to have a condition independent of that,
  we must have term-by-term equality for the above sum. That is, we must have (canceling out the common factors)
  \beq 
  e^{i(\ga_{\be}(n) - \ga_{\al}(n))} = \frac{\balb}{[\balb]^{*}}.
  \label{Cond_gamma}
  \eeq 
  \noi At this stage, it is clear that this will be a feasible condition with $\gamma_{\al}$s independent of $n,$ so we drop $n-$dependence of $\gamma_{\al}$s. \\
  
  Now, 
  \beq
  \balb = \la x_{\al} |U_{z}|x_{\be}\ra.
  \label{balb_Expr}
  \eeq 
  \noi Let us denote by $n_{\al\be},$ the number of spins needed to be flipped to generate $\xal$ from $\xbet$. Then from the Taylor expansion of $U_{z},$  it can be shown that when $n_{\al\be}$ is odd, then $\balb$ is imaginary hence $\frac{\balb}{[\balb]^{*}} = -1$ and 
  when $n_{\al\be}$ is even, then $\balb$ is real, hence $\frac{\balb}{[\balb]^{*}} = +1.$ Thus, we can write
  \beq 
   \frac{\balb}{[\balb]^{*}} =  e^{i\pi n_{\al\be}} = e^{i(n_{\al} - n_{\be})}.
  \eeq 
  The proof of the last step of the above equation (the assignment) follows this part of the proof. This is equivalent for the present purpose to the
  individual assignments of $\gamma_{\al}$s as
  \beq
  \ga_{\al} = n_{\al}\pi, 
  \label{gamma_form}
  \eeq 
  \noi where $n_{\al} =$ the number of spins required to be flipped to create $\xal$
  from some pre-assigned reference state $|R\ra$ (e.g., all-up $x-$basis state). 
  This gives the concrete form of the necessary starting condition for the
  induction proof (Eq.~\ref{nth_Cond},~\ref{n+1_th_Cond}). The condition now reads: if after $n-$th cycles of evolution by $U$ and $\tilde{U}$ the coefficients satisfy the condition
  \beq 
  [a_{\al}^{(n)}]^{*} = e^{-i\pi n_{\be}}\tilde{a}_{\al}^{(n)},
  \label{nth_Cond_Final}
  \eeq 
  \noi then the same must hold after $(n+1)-$th step, i.e, we must have
  \beq 
  [a_{\al}^{(n+1)}]^{*} = e^{-i\pi n_{\be}}\tilde{a}_{\al}^{(n)},
  \label{nth_Cond_Final}
  \eeq 
  This statement can be proved immediately following the steps sketched above. 
  Now it is easy to see that this satisfied if we take $|\psi(0)\ra = \xal$ (any
   $x-$basis state). This completes the proof.
  \\
  
  \noi 
 {\it Proof of the Assignment:} Here we prove that 
 $e^{i\pi n_{\al\be}} = e^{i\pi(n_{\al} - n_{\be})}.$ \\
  
  \noi Let  $n_{\al} =$ the number of spins required to be flipped to create $\xal$
  from the all-up $x-$basis state, and $n_{\be}$ is defined likewise. \\
  
  \noi Let $n_{c} = $ the number of common spins that are flipped in both $\xal$ and $\xbet$ compared to the reference state $|R\ra$. \\
  
  \noi Then $\tilde{n}_{\al} = n_{\al} - n_{c} = $ No. of spins that are flipped in $\xal$ (w.r.t $|R\ra$) but not flipped in $\xbet.$ \\
  
  \noi Similarly, $\tilde{n}_{\be} = n_{\be} - n_{c} = $No. of spins that are flipped in $\xbet$ but not flipped in $\xal.$ \\

\noi
Thus $n_{\al\be}  = \tilde{n}_{\al} + \tilde{n}_{\be}.$ \\

\noi
Now, all we care about is the parity (whether odd/even) of $ P(n_{\al\be}).$ 
But from above we have 
$$
n_{\al\be} = \tilde{n}_{\al} + \tilde{n}_{\be} = n_{\al} + n_{\be} - 2n_{c}.
$$
\noi Since $2n_{c}$ is always even, we hence have
\bea 
P(n_{\al\be}) = P(\tilde{n}_{\al} + \tilde{n}_{\be}) &=& P(n_{\al} + n_{\be}) \non 
\eea 
\noi But it for any integers $n_{\al}$ and $n_{\be},$ it is evident that 
$P(n_{\al} + n_{\be}) =  P(n_{\al} - n_{\be}).$ Thus we finally have
\bea 
P(n_{\al\be}) &=& P(n_{\al} - n_{\be}) \non \\
\implies e^{i\pi n_{\al\be}} &=& e^{i\pi(n_{\al} - n_{\be})} ~ {\rm (proved)}.
\eea


\bibliographystyle{apsrev4-2}
\bibliography{DTC_reference}

\end{document}